\documentclass[a4paper,onecolumn,11pt]{quantumarticle}
\pdfoutput=1
\usepackage[utf8]{inputenc}
\usepackage[english]{babel}
\usepackage[T1]{fontenc}
\usepackage{amsmath,amssymb,latexsym}
\usepackage{bbm}
\usepackage{hyperref}

\usepackage{tikz}
\usepackage{lipsum}
\usepackage{svg}

\PassOptionsToPackage{compress}{natbib}

\usepackage[numbers,sort&compress]{natbib}

\renewcommand{\tilde}{\widetilde}
\renewcommand{\bar}{\overline}

\def \oh {\frac{1}{2}}

\def \al {\alpha}

\def \ep {\epsilon}

\def \Ec {\mathcal{E}}

\def \Oc {\mathcal{O}}

\def \Nc {\mathcal{N}}

\def \pr {\partial}
\def \ra {\rightarrow}

\def \oh {\cfrac{1}{2}}

\def \beq { \begin{equation}}
\def \eeq {\end{equation}}

\def\bal#1\eal{\begin{align}#1\end{align}}

\DeclareMathOperator*{\Tr}{Tr}

\newcommand\const{\operatorname{const}}

\def \l {\left}
\def \r {\right}

\def \bra {\langle}
\def \ket {\rangle}

\def \Qh {\hat{Q}}
\usepackage[numbers]{natbib}

\DeclareUnicodeCharacter{1EF3}{\^{i}}

\begin{document}

\title{Observable-projected ensembles}

\author{Alexey~Milekhin$^*$}
\affiliation{Institute for Quantum Information and Matter, California Institute of Technology, Pasadena, CA 91125, USA}
\author{Sara~Murciano$^*$}
\footnotetext{$^*$Equal contribution.}
\affiliation{Institute for Quantum Information and Matter, California Institute of Technology, Pasadena, CA 91125, USA}
\affiliation{Walter Burke Institute for Theoretical Physics,  California Institute of Technology,  Pasadena, CA 91125, USA}
\affiliation{Department of Physics, California Institute of Technology, Pasadena, CA 91125, USA}
\maketitle

\begin{abstract}
 Measurements in many-body quantum systems can generate non-trivial phenomena, such as preparation of long-range entangled states, dynamical phase transitions, or measurement-altered criticality. Here, we introduce a new measurement scheme that produces an ensemble of mixed states in a subsystem, obtained by measuring a local Hermitian observable on part of its complement. We refer to this as the \textit{observable-projected ensemble}. Unlike standard projected ensembles—where pure states are generated by projective measurements on the complement—our approach involves projective partial measurements of specific observables. This setup has two main advantages: theoretically, it is amenable to analytical computations, especially within conformal field theories. Experimentally, it requires only a linear number of measurements, rather than an exponential one, to probe the properties of the ensemble. As a first step in exploring the observable-projected ensemble, we investigate its entanglement properties in conformal field theory and perform a detailed analysis of the free compact boson.
\end{abstract}

\section{Introduction}

In recent years quantum dynamics involving measurements has drawn a lot of attention in the literature, driven by multiple factors. First of all,
any interaction of a quantum computer with a classical observer will include a measurement. With the ever-growing number of quantum-computing experiments, it is hard to overestimate the importance of this. For instance, performing measurements on a subsystem of a resource state can be used to prepare non-trivial quantum states with
long-range correlations, and this is the basis of measurement-based quantum computation~\cite{nat21,Lu22,nat22,bravyi_2022,leeji2022,Zhu22,piroli2021}. On a more fundamental level, the dynamics with measurement can exhibit a rich phase structure~\cite{Li:2018mcv, Li:2019zju, Skinner:2018tjl,Szyniszewski:2019pfo, Szyniszewski:2020fwl,Choi:2019nhg,Vijay:2020yuk,Li_2021, Altland:2021bqa,Milekhin:2022bzx,Yoshida:2022srg}, which in some cases can be explained by an emergent quantum error correction in an ensemble of states~\cite{Fan:2020sau}. Moreover, measurements can be used to alter the pristine entanglement properties of critical states~\cite{AltmanMeasurementLL,garratt24,Weinstein23,entropyLuttinger,usmeasurementaltered,Paviglianiti2023,patil2024}, as well as to uncover the intricate inner structure of pure states, which is the subject of this paper.

%Suppose we have a fixed pure state $|\Psi \ket$. The total system consists of a region $A$ and its compliment $B=\bar{A}$. If the state $|\Psi\ket$ is the result of a long-time ergodic evolution, then for large enough $B$ we should expect to find $A$ in an appropriately defined canonical ensemble after tracing out $B$:
%\beq
%\label{eq:trace_out}
%\Tr_B | \Psi \ket \bra \Psi | \approx e^{-\beta H_A}/\Tr e^{-\beta H_A}.
%\eeq

{\bf Projected ensembles}: Given a system $A\cup \bar{A}$, if the state $|\Psi\rangle$ is the result of a unitary evolution, at late times, the local stationary behavior is described by a statistical ensemble, corresponding to a thermal or generalized Gibbs ensemble for chaotic or integrable systems, respectively \cite{therm2,therm3}. This implies that the reduced density matrix $\rho_A=\mathrm{Tr}_{\bar{A}} | \Psi \rangle \langle \Psi |$ is the main conceptual tool to understand how and in which sense an isolated quantum system can be described by a mixed state at large times.
A natural definition that arises from the reduced density matrix is a family of entanglement measures, the R\'enyi entropies, which are given by 
\begin{equation}\label{eq:defrenyi}
    S_A^{(n)}=\frac{1}{1-n}\log \Tr [\rho_A^n], \qquad S_A^{(1)}=-\Tr[\rho_A\log\rho_A].
\end{equation}
The right part of the equation above is the definition of the von Neumann entanglement entropy, one of the most successful entanglement measures, which can be derived by doing an analytic continuation in $n$ of $S_A^{(n)}$ and then taking the limit $n\to 1$~\cite{cc-04}.

The question of whether the thermalization is even approximately true or not, and how to properly describe it, has been the focus of quantum statistical mechanics. The above statement, however, raises questions about the coarse-grained nature of $A$: if we average over the state of $\bar{A}$, what is the most appropriate way to describe the state of $A$? Modern quantum experiments have put forward a new perspective, in which one can consider a \textit{projected ensemble} of pure states corresponding to different microscopic states of $\bar{A}$: given a set of states $|z_i \ket$ in $\bar{A}$, we define the projected ensemble $\Ec$ of states of $A$ as
\beq\label{eq:proj_ens}
\Ec =  \{ p_i, \ | \Psi_i \ket =   p_{i}^{-1/2} \bra z_i |\Psi \ket \},
\eeq
where $p_i = |\bra z_i | \Psi \ket|^2$ are the corresponding probabilities of finding $\bar{A}$ in the state $| z_i \ket$ . Then, for a complete set of states of $\bar{A}$, the \textit{average} state $|\Psi_i\ket$ reduces to $\rho_A$, since
\beq
\rho_1 = \mathbb{E}_\Ec | \Psi_i \ket \bra \Psi_i | = \sum_i p_i 
|\Psi_i \ket \bra \Psi_i | = 
 \sum_i \bra z_i | \Psi \ket \bra \Psi | z_i \ket = 
\Tr_{\bar{A}} | \Psi \ket \bra \Psi|.
\eeq
However, a similar operation is far from trivial if we consider the higher moments of $\mathcal{E}$, i.e. averaging $(|\Psi_i\ket\bra\Psi_i|)^{\otimes k}$. 
Indeed, one main feature of the projected ensemble is its convergence to a uniform distribution over the set of pure states in $A$, the \textit{Haar ensemble},
for chaotic dynamics and infinite-temperature initial states. This phenomenon has been dubbed deep
thermalization and it represents a form of equilibration in quantum many-body systems stronger than the
regular thermalization we mentioned above, which only constrains the expectation values of observables over the ensemble of the stationary state~\cite{cotler23,choi23}.
It has been also proven in different setups like free fermionic systems~\cite{piroli23}, deep random circuits~\cite{cotler23,chan2024}, dual-unitary models~\cite{ho22,Ippoliti2022,Claeys2022,Ippoliti23d,ho23}, under charge-conserving quantum dynamics~\cite{chang24}, and then also extended to finite temperature cases~\cite{federica}. 
Nevertheless, beyond dynamical settings, the ground state properties of the ensemble~\eqref{eq:proj_ens} have not been further studied so far. To the best of our knowledge, the main features one can derive about $\mathcal{E}$ after a complete set of measurements is that, if $|\Psi\rangle$ is an infinite temperature state, then $\mathcal{E}$ is the Haar ensemble, while at finite temperature $\mathcal{E}$ is the Scrooge ensemble (a deformation of the Haar random ensemble~\cite{jrw-94}) corresponding to the appropriate thermal reduced density matrix. The goal of this paper is to investigate more projected ensembles originating from the ground states.

\paragraph{Entanglement after partial projective measurements:} Despite studying the static properties of $\mathcal{E}$ is challenging, focusing on specific measurement outcomes is more accessible for analytical computations. For instance, many interesting quantum many-body systems in 1+1 dimensions possess quantum critical points whose ground state can be described by conformal field theories (CFT) at long distances, and there has been a moderate progress in describing post-measurement states $|\Psi_i\rangle$ of CFT using the boundary CFT (BCFT) description~\cite{Rajabpour_2015,Rajabpour_2016}. Unfortunately, these are applicable only by choosing the basis and the outcome $|z_i\rangle$ in a way that
the induced boundary condition on $B$ is still conformally invariant. For instance, we can measure the transverse magnetization in the XX spin chain (aka free fermionic model) and post-select a measurement outcome. In the absence of a magnetic field, the most likely outcome is an antiferromagnetic string. It
is expected that this case leads to the Dirichlet boundary condition in the bosonization language and so it is related
to the BCFT.
More generally, the main result about specific measurement outcomes is that, after partial projective measurements, if $B$ is a subsystem of length $s$ in an infinite system, $A$ is a region of length $\ell$ adjacent to $B$, and $\epsilon$ is a UV cutoff, then the R\'enyi entanglement entropies behave, at leading order, as~\cite{Rajabpour_2015}
\begin{equation}\label{eq:rajabpour}
    S^{(n)}_A=\frac{c}{12}\frac{n+1}{n}\log \frac{\ell(\ell+s)}{s\epsilon}+\log b_0 %+\frac{b_1^2{b_0^2}\left(\frac{s\epsilon}{2\ell(\ell+s)}\right)^{2\Delta_1/n},
\end{equation}
where $c$ is the central charge of the CFT and $\log b_0$ is the Affleck-Ludwig boundary term. This result has been generalized to several different geometries, including cases where $B$ is not a simply connected region. We can recognize in Eq.~\eqref{eq:rajabpour} the leading order behavior of the entanglement entropy in the presence of a boundary.
However, not all the measurement outcomes lead to conformal invariant boundary conditions. For instance, if we post-select the ferromagnetic string, it does not lead to a BCFT, because it already breaks the $U(1)$ symmetry~\cite{Stephan_2014,Stephan_20142}.
The challenging part of this problem is not only the extension to the non-conformal invariant setups, but also the computations of quantities like the \textit{localizable entanglement}~\cite{localizable} or the \textit{measurement induced entanglement} (MIE)~\cite{Lin2023probingsign}. The latter is defined for a tripartite geometry, $A_1,A_2$ and $B$, where we projectively measure $B$ and the MIE between $A_1$ and $A_2$ is defined as 
\begin{equation}\label{eq:defMIE}
    \mathrm{MIE}(A_1:A_2)=\sum_i p_i S^{(1)}_{A_1}(|\Psi_i\rangle) \qquad S^{(1)}_{A_1}(|\Psi_i\rangle)=-\mathrm{Tr}\rho_{A_1}\log  \rho_{A_1}.
\end{equation}
Here $i$ parameterize the measurement outcomes with probabilities $p_i$ and $\rho_{A_1}=\mathrm{Tr}_{A_2}|\Psi_i\rangle \langle \Psi_i|$. Since both the MIE and the localizable entanglement involve a sum of all possible measurement outcomes, it is hard to analytically predict their behavior, since we do not have a closed formula for each $S^{(1)}_{A_1}$ in a field theory setup. However, the MIE can be used to gain insights about the initial state $| \Psi \ket$, for example, the sign structure of stabilizer states~\cite{Lin2023probingsign}, or whether measurements can generate entanglement between distant parties without the need for direct interaction. 

\paragraph{Main results:} Even though a lot of effort has been done to understand the out-of-equilibrium properties of projected ensembles, finding result in the ground state of critical theories is far from being trivial due to the main challenges we summarized above. 
\begin{figure}[ht!]
\centering
    {\includegraphics[width=0.6\textwidth]{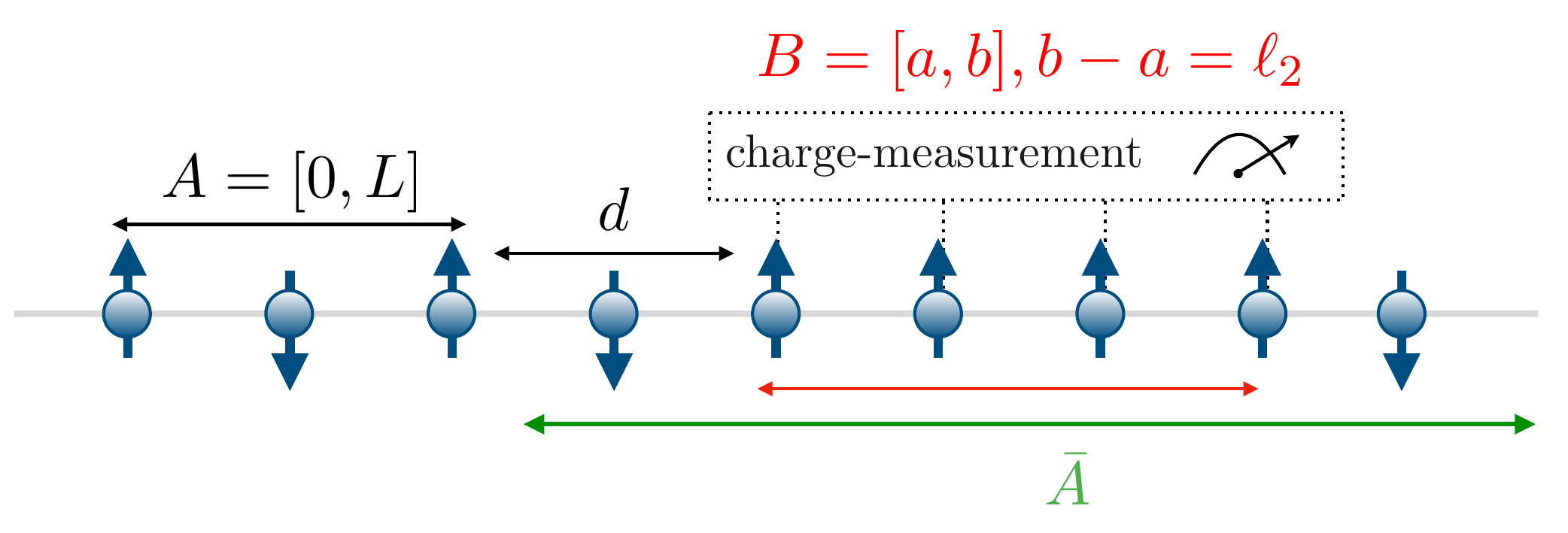}}
     \caption{The geometry we consider in this manuscript is the following: we measure an observable (such as the charge operator) in a region $B$ and we study the properties of the reduced density matrix $\rho_A$ after this operation. }
     \label{fig:geometry}
\end{figure}
\textit{The focus of this paper is exactly to approach the structure of a projected ensemble using a field-theoretic framework.}
Fortunately, oftentimes we are not interested in the state of the system after a complete projection of its subsystem $B$. If we consider experiments involving $M$ measurements, having sets of outcomes $m$, we can characterize the post-measurement density
matrix $\rho_m$ by computing an expectation value. To do that, we must prepare
the same state $\rho_m$ multiple times. The problem is that the probability to observe $m$, $p_m$, is in general, exponentially small in
the system size of $B$. Thus, we need $M\gg 1/p_m$ to observe $m$ more than once. This means that we need exponentially many (in the size of $B$) outcomes, and sampling from this ensemble is hard. In contrast, measuring a certain observable inside $B$ will lead to a number of outcomes that scale at most linearly with the size. 
This is why in this paper we will introduce a smaller ensemble which corresponds to projecting on a particular value of an extensive observable, rather than a complete projective measurement of $B$. As we will see, such an ensemble has an elegant field-theoretic description.
Specifically, we will consider a local Hermitian operator $\Oc$ and ask about the ensemble of states having a definite value $q$ after measuring $\sum_{j\in B}\Oc_j$, or its continuum counterpart $\int_B dx \ \Oc$ (in taking the continuum limit of the observable $O$, we omit explicit lattice regularization
\footnote{ One very important subtlety is whether this delta-function really represents a projection operator. 
In order to see that it is indeed true, let us assume for simplicity that we have a discrete integer spectrum and use an integral representation for the delta-function:
$$
\frac{1}{2\pi} \int_0^{2\pi} d\alpha \ e^{i \alpha (q- \sum_{j \in B} \Oc_j)}
$$
If we act with this expression on a state $| S \ket$ which is an eigenstate of $\sum_{j \in B} \Oc$ with eigenvalue $S$, we get
$\frac{1}{2\pi} \int_0^{2\pi} d\alpha  \ e^{i \alpha (q-S)} | S \ket \propto \delta_{S,q} |S \ket$.
The last delta $\delta_{S,q}$ is the standard discrete Kronecker delta. Hence our construction indeed acts as a projection operator.
Notice that this is a non-trivial check since all this argument would not have worked if we had another operator $V$ multiplying the expression in the delta-function: 
$\delta(V(q- \sum_{j \in B} \Oc_j))$
We thank anonymous referee for raising this question.
}
)
\beq\label{eq:mathEO}
\Ec_\Oc = \left\{ p_q , \  \Pi^B_q | \Psi \ket = \delta \l( q - \int_B dx \ \Oc \r) |\Psi \ket \right\},
\eeq
being $p_q$ the probability of finding $q$ as an outcome of the measurement of our observable.
This setup is closely related to the projected ensembles introduced at the beginning but differs from it in three key aspects. First, we are obtaining very limited information about $B$ (the value of $\int_B dx \ \Oc$). Second, the state we get after the partial projection of $\int_B dx \ \Oc$ and tracing over $\bar{A}$ is not pure anymore, contrarily to what happens after a standard projective measurement in the limit in which $B=\bar{A}$. Beyond Ref.~\cite{jrw-94}, we are not aware of settings where $\mathcal{E}$ consists of a set of mixed, rather than pure, states at equilibrium. Such a scenario is actually more realistic than the one with pure states, and, indeed, we can think of $\mathcal{E}_{\mathcal{O}}$ as an ensemble of mixed states.
Third, we ask questions not about the complement of $B$, but rather about a smaller subsystem which we will keep calling $A$. Given these differences, we dub $\mathcal{E}_{\mathcal{O}}$ an \textit{observable-projected ensemble} (see Fig.~\ref{fig:geometry} for a visualization of the geometry we are considering).

In this paper, we discuss the general properties of $\Ec_\Oc$ and we evaluate the entanglement entropy $S_A$ of the region $A$, for both fixed post-measurement states $\Pi^B_q | \Psi \ket$ and averaged over the measurement outcomes:

\beq\label{eq:SIE}
{\rm MIE}(A: \bar{A\cup B})=\sum_q p_q S^{(n)}_A( \Pi^B_q | \Psi \ket),
\eeq
where $S^{(n)}_A$ are the R\'enyi entropies defined in Eq.~\eqref{eq:defrenyi}.
This quantity is again a MIE, according to the definition in Eq.~\eqref{eq:defMIE}, where now $A_1=A, A_2=\bar{A\cup B}$ is an infinite subsystem and $S_A^{(n)}$ refers to the entanglement entropy of $A$ in the post-measurement state.
%has been previously called \textit{measurement-induced entanglement} (MIE)~\cite{another huge set of citations here}.
%Like $\rho_k$ and the related deep-thermalization (\ref{eq:k_moment}), 

If $\Oc$ is the density of a conserved charge, then the quantity we are studying is closely related to the symmetry-resolved entanglement (see~\cite{review} for a review). Given a subsystem $A$, the reduced density matrix of $A$ often consists of blocks corresponding to different charge sectors and the
symmetry-resolved entanglement is the entanglement in each sector. Here, the main difference is that we project into a charge sector of $B$ and then we study the entanglement entropy between $A$ and its complement (including $B$). 
%Hence, it would be natural to call this quantity \textit{symmetry-projected entanglement}, even though we do not limit ourselves to setups where the total charge is conserved.

One technical challenge in defining the observable-projected ensemble $\Ec_\Oc$
is whether the related quantities we compute in the field-theoretic approximation are universal (UV-insensitive) or not. Renormalizable quantum field theories (QFTs) are distinguished by the fact that all UV divergences in all correlation functions can be removed after fixing a small subset of data (e.g. lower-point correlation functions at a given kinematics).
It is not entirely obvious if the moments of $\Ec_\Oc$ have a similar property. Technically, the problem comes from the fact that $B$ is a finite interval. However, we will argue that if $\Oc$ has low conformal dimension, it is possible to obtain universal answers about the moments of $\Ec_\Oc$.

After the introduction of the observable-projected ensembles $\mathcal{E}_{\mathcal{O}}$ as a new theoretical tool suitable for modern quantum experiments, which can generate and investigate ensembles of states labeled by specific measurements, we summarize here the key theoretical findings about $\mathcal{E}_{\mathcal{O}}$ that we present in this manuscript. 
The first setup in which we develop our analysis is when the measured observable is the charge of the system, such as the generator of a $U(1)$ symmetry. This is the main content of section~\ref{sec:chargeproj}. We show how a potential experimental setup involving randomized measurements can be used to study the properties of this ensemble.
This setting is also suitable for a field theory study, since a prototypical system with a global $U(1)$ symmetry can be described by a simple CFT, a free compact boson with central charge $c=1$, as we show in section~\ref{sec:freecompact}. In this case, we perform partial measurements of the number operator in a subsystem and, if we focus on a specific charge sector, it turns out that the dependence on the measurement outcome simplifies and Eq.~\eqref{eq:SIE} reduces, at leading order, to the total entanglement entropy with a subleading contribution depending on the geometry of the charge-projected region. As a consequence of this result, one may conclude that measuring the charge operator does not have drastic changes on the initial state $|\Psi\ket$. For this specific theory, we can also infer the upper bound to the amount of accessible information we can extract by varying the size of the measured region.

This conclusion naturally raises the question about what happens measuring $\int_B dx\, \Oc$ for an observable $\Oc$ different from the charge density. To address this problem in CFT, we consider the projection of some other Hermitian local operator, which is a primary field with Gaussian correlators. In this setup, we find that each $S^{(n)}_A( \Pi^B_q | \Psi \ket)$ becomes a non-trivial function of both $q$ and the size of the projected region, as we show in section~\ref{sec:others}. As a consequence, the dependence of the MIE \eqref{eq:SIE} on the geometry we consider becomes more intricate. Moreover, in our field theory setup, this expression is divergent and non-universal, and we aim to find a finite expression for the MIE. By using different weights for each measurement outcome, rather than simply $p_q$ as in Eq.~\eqref{eq:SIE}, we can also construct universal quantities for the observable-projected ensemble in section~\ref{sec:universal}. We extend our results to non-Gaussian CFTs, even though finding UV-finite expressions for generic theories is more challenging and requires more constraints on the operator we can measure. Finally, for Gaussian states, we develop a strategy to compute numerically the entanglement properties of the observable-projected ensembles in section~\ref{sec:num_checks}.

\section{Charge-projected ensemble}\label{sec:chargeproj}

As a first and more intuitive example of an observable-projected ensemble, we start from a system with a global $U(1)$ symmetry, i.e. the number of particles. If we denote by $Q$ the total charge and $|\Psi\rangle$ is its eigenstate, then the total density matrix $\rho =|\Psi\rangle\langle \Psi|$ satisfies $[\rho,Q]=0$. We consider a geometry in which $A$ and $B$ are not complementary parts (see Fig.~\ref{fig:geometry}), so we denote by $\bar{A}$ and $\bar{A\cup B}$ all the region outside $A$ and $A\cup B$, respectively, and the total charge splits as $Q=Q_A+Q_B+Q_{\bar{A\cup B}}$. The Hilbert space associated to the subsystem $A$ ($\bar{A}$) is $\mathcal{H}_A$ ($\mathcal{H}_{\bar{A}}$).
If now we act with a unitary operator $U_B$ only in the subregion $B$, we do not modify the entanglement between $A$ and $\bar{A}$, because $\rho_A=\Tr_{\bar{A}}[U_B\rho U^{\dagger}_B]=\Tr_{\bar{A}}\rho$, due to the cyclicity of the trace. However, if we perform a more drastic operation in $B$, like the projection into a given charge sector of $Q_B$, then the result is not trivial. In principle, we want to study
\beq
\Ec_Q = \left\{ p_q , \  \Pi^B_q | \Psi \ket \right\},
\eeq
where $q$ is an eigenvalue of $Q_B$ and $\Pi^B_q$ is the projection operator into the $q$-charge sector. At this level, $\Ec_Q$
is an ensemble of pure states living in $\mathcal{H}_{A}\otimes \mathcal{H}_{\bar{A}}$. However, if we consider a full projection in the subsystem $B$, as in Eq.~\eqref{eq:proj_ens}, then the resulting state is defined only on $\mathcal{H}_A$. Thus, since we are interested in a framework similar to the full projected ensemble but after measuring a certain observable inside $B$, we focus our attention on the following ensemble of mixed states
\beq\label{eq:rhoAq}
\mathcal{E}_{Q}^{\rm mixed} = \left\{ p_q,\rho_{A,q}=\Tr_{\bar{A}} \Pi^B_q \rho \Pi^B_q \right\}.
\eeq

To compute the MIE in Eq.~\eqref{eq:SIE}, which is our main goal, we can exploit the Fourier representation of $\Pi^B_q$
%where we can write $\Pi^B_q$ as
\beq
\Pi^B_q = \frac{1}{2\pi}\int d \alpha \  e^{i \alpha Q_B-i \alpha q },
\eeq
such that we rewrite Eq.~\eqref{eq:rhoAq} as
\beq
\label{eq:rhoAq_simp}
\rho_{A,q} = \frac{1}{2 \pi \Nc} \Tr_{\bar{A}} \int d\al_1 d\al_2 \ e^{-i \alpha_2 q} e^{i \alpha_1 q} e^{i \alpha_2 Q_B} \rho e^{- i \alpha_1 Q_B}=\frac{1}{\Nc}  \int d\gamma  \ e^{-i \gamma q} \Tr_{\bar{A}}( e^{i \gamma Q_B}\rho ),
\eeq
after the change of variables $\gamma=\alpha_1-\alpha_2$. Here $\Nc$ is the normalization factor 
\beq
\Nc = \int d\alpha \ e^{-i q \alpha } \Tr ( e^{i Q_B  \alpha} \rho ),
\eeq
that ensures that $\mathrm{Tr}\rho_{A,q}=1$. Let us remark that $\mathrm{Tr}$ simply denotes the trace over the full system. 
If we are interested in the replicated version of the problem, as in the computation of the R\'enyi entropies in Eq.~\eqref{eq:defrenyi}, the object we need to compute is
%\beq\label{eq:tr1}
%\Tr \rho_{A,q}^n = \frac{1}{\Nc^n} \int d\alpha^{(1)}_1 d\alpha^{(1)}_2 \dots d\alpha^{(n)}_1 d\alpha^{(n)}_2e^{-iq\sum_k (\alpha^{(k+1)}_{2}-\alpha^{(k)}_1)} \Tr
%\prod_{k=1}^{n} [\rho e^{i\Qh_B (\alpha^{(k+1)}_{2}-\alpha^{(k)}_1)}],
%\eeq
%with $\alpha_2^{(n+1)}\equiv \alpha^{(1)}_2$.
%By doing a simple change of variables
%\beq
%\gamma_k = \alpha^{(k+1)}_2 - \alpha^{(k)}_1,
%\eeq
%we can rewrite Eq.~\eqref{eq:tr1} as 
\beq\label{eq:tr2}
\Tr \rho_{A,q}^n = \frac{1}{\Nc^n} \int d\gamma_1 d\gamma_2 \dots d\gamma_n e^{-iq\sum_k \gamma_k} \Tr
\prod_{k=1}^{n} \Tr_{\bar{A}}[\rho e^{i\Qh_B \gamma_k}].
\eeq
By using Eq.~\eqref{eq:tr2}, we provide a prescription to analyze the properties of the observable-projected ensemble when the operator is the generator of the $U(1)$ symmetry. We remark that this prescription is valid to investigate $\mathcal{E}_Q$ both on the lattice and at the field theory level, even though the primary focus of this manuscript is the study of this quantity in CFTs. 

As an application of these result, we can study the following protocol. One party, Bob performs the charge measurement on $B$. The other party, Alice, wants to determine Bob's measurement outcome. However, she only has access to the subregion $A$. In other words, we want to understand how much \textit{accessible information}~\cite{Nielsen:2010qu,jozsa94} does the ensemble $\rho_{A,q}$ have. For this purpose, we use the \textit{Holevo-$\chi$ quantity}~\cite{Holevo}, which provides an upper bound for the accessible information and it is given by $\chi(\mathcal{E}_\Oc ) \equiv S_A^{(1)} (\rho_{A} ) - \sum_q p_q S^{(1)}_A( \Pi^B_q | \Psi \ket)$. 

Before showing how to compute Eq.~\eqref{eq:tr2} in a specific theory, we will outline how the charge-projected ensembles can be experimentally probed using measurements and additional classical computational steps.

\subsection{Randomized measurements as an experimental probe}

We can ask whether it is possible to design an experimental implementation to study the properties of our charge-projected ensembles. We first observe that, if we can prepare a non-trivial state on $N$ sites, then measuring the possible values of the charge is much easier than considering all the possible projective measurements. Indeed, the first set of measurements scales linearly with the system size ($O(N)$), while the second set grows exponentially ($O(e^{N})$). Therefore, studying the reduced density matrix in different charge sectors $q$ is not prohibitive. 

For this purpose, one can employ the randomized measurement toolbox, which is particularly efficient in estimating quantum state properties expressible as polynomial functions of a density matrix (see~\cite{elben_review} for a review). First, we need an unbiased estimator of the reduced density matrix after a projection of the charge, $\rho_{A,q}$, called classical shadow. A way to measure the charge in quantum circuits can be found in~\cite{piroli2024}. This requires a number of ancillae and a circuit depth scaling logarithmically with the subsystem size. Morally, in this approach, one prepares an auxiliary particle in a position eigenstate $|x=0 \ket$, applies the operator $e^{i Q_B p}$, where $p$ is the momentum conjugate to $x$ and then measures the new particle position.

We can get the classical shadow by applying a random unitary transformation $U$ to the quantum state $\rho_{A,q}$, where $U$ is typically chosen from a unitary 2-design, such as the Clifford group or the Haar measure over the unitary group. We then measure the transformed state $\rho^q_U=U^{\dagger}\rho_{A,q} U$ in the computational basis (e.g. the number operator in this case) to obtain an outcome $|s^q\rangle$, where $s^q$ is a bit string representing the measurement outcome.
%We post-select only for the measurement outcome in a given charge sector, $q$, and we also collect the probability of getting the configuration $s^q$ with charge $q$, which we denote by $p_{s^q}$. 
For each measurement outcome, we can construct an estimator of the original state $\rho_{A,q}$, as
\begin{equation}
    \hat{\rho}^q_U=d U^{\dagger}|s^q\rangle \langle s^q|U-(d-1)\mathbbm{1}/d, 
\end{equation}
where $d$ is the Hilbert space dimension. By repeating the above steps for multiple random unitaries $U_i$
and measurement outcomes $|s^q_i\rangle $, that we denote by $M$, the classical shadow of the quantum state $\rho_{A,q}$ is the average of the individual estimators 
\begin{equation}\label{eq:estimator}
     \hat{\rho}_{A,q}=\frac{1}{M}\sum_{i=1}^M\hat{\rho}^q_{U_i}.
\end{equation}
The above estimator is unbiased, in the sense that $\mathbbm{E}[\hat{\rho}_{A,q}]=\rho_{A,q}$, and the expectation value is taken over randomized measurements. There are different options that one might employ to boost the convergence to a better estimator, for instance by combining independent realizations of the classical shadow $\hat{\rho}^q_{U_i}$
\cite{Neven2021,Vitale2022}. Even though we will not perform a thorough statistical analysis here, we believe that this approach can measure the properties of $\rho_{A,q}$ and of $\mathcal{E}_{Q}^{\rm mixed}$ in various NISQ
platforms up to moderate partition sizes, especially because the ensemble we consider only requires a linear number of measurements of $Q_B$, rather than exponential. However, even though postselection is not a problem in this case, since $\rho_{A,q}$ is a mixed state, its moments (e.g. the purity) are exponentially small in system size, and so measuring them is still a hard task.  

Finally, we observe that, rather than doing a pre-selection on the measurement charge, one might also construct an unbiased estimator for $\rho_{AB}$, i.e. the reduced density matrix on $A\cup B$, and then analytically project into the eigenspace of $Q_B$ with eigenvalue $q$. A similar approach has already been used to measure the symmetry-resolved entanglement or related quantities~\cite{Vitale2022,Neven2021,rath23,joshi24}. However, since we are interested in measuring the entanglement, the construction for the estimator of $\rho_{AB}$ is more expensive as the Hilbert space dimension would be larger than only $\rho_{A,q}$.

\section{A case study: the free compact boson}\label{sec:freecompact}
To give a more concrete estimate of the charge-projective ensemble, let us focus on a compact boson, which is the CFT of the Luttinger liquid, described by the action 
\begin{equation}
    S=\frac{1}{2}\displaystyle \int d^2z(\partial_{\mu}\varphi)^2.
\end{equation}
The target space of the real field $\varphi$ is compactified on a circle of radius $R$, so the action is invariant under the transformation $\varphi\to \varphi+\alpha$
which, due to the compact nature of $\varphi$, realizes a $U(1)$ global symmetry. The compactification radius $R$ is proportional to the Luttinger parameter $K$ ($R\propto K^{-1/2}$).
We can further fix the geometry to be $A=[0,L]$, $B=[a,b]$, with $a,b>L, b-a=\ell_2$ (see Fig.~\ref{fig:geometry}). The first object we want to compute is $ \Tr
\prod_{k=1}^{n} \Tr_{\bar{A}}[\rho e^{iQ_B \gamma_k}]$, a quantity similar to the charged moments discussed in~\cite{Ares_2023} to evaluate how much a symmetry is broken in a subsystem. Indeed, we stress that even though $[\rho,Q]=0$, once we restrict the charge to the subsystem $B$, $[\rho,Q_B]\neq 0$, and we need to be careful in preserving the order of the non-commuting operators in Eq.~\eqref{eq:tr2}. 

\begin{figure}
    \centering
    \includegraphics[scale=0.8]{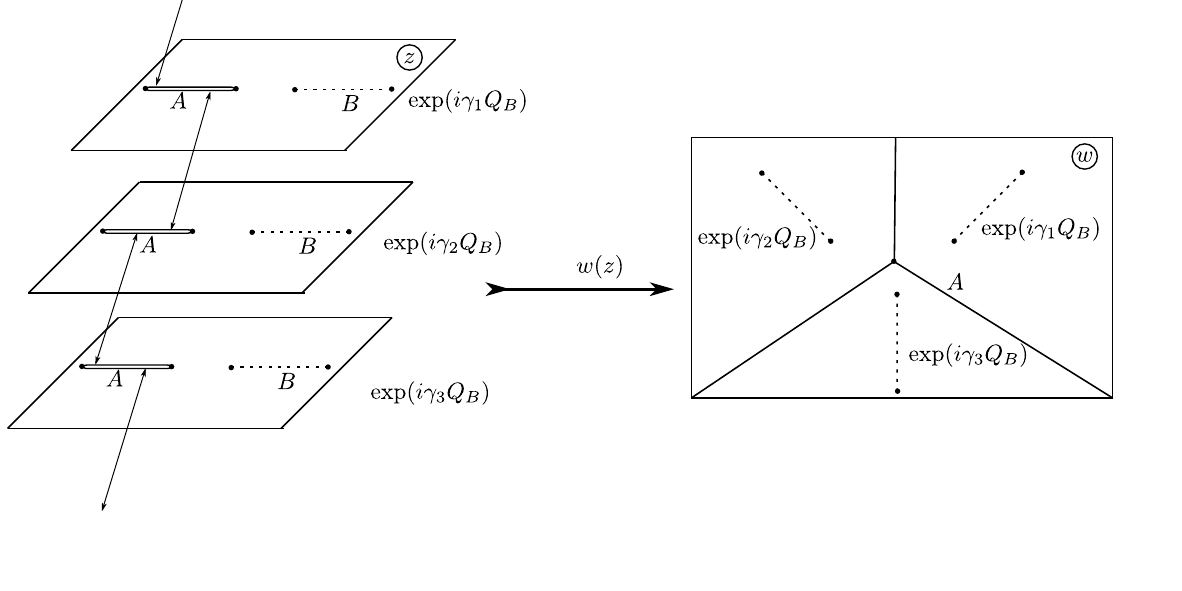}
    \caption{Mapping the replica geometry $\mathcal{R}_n$ to a single complex plane using Eq.~(\ref{eq:z_to_w}).}
    \label{fig:cft_sheets}
\end{figure}

The charged moments are defined on the $n$-sheeted Riemann surface $\mathcal{R}_n$ shown in the left panel of Fig.~\ref{fig:cft_sheets} and parametrized by the coordinate $z$.
In order to evaluate them, we first map $\mathcal{R}_n$ to the complex plane $\mathbb{C}$ in Fig.~\ref{fig:cft_sheets} via the
uniformization map 
%Mapping replica manifold parametrized by $z$ (Figure~\ref{fig:cft_sheets} (a) ) to a sphere parametrized by $w$ (Figure~\ref{fig:cft_sheets} (b) ):
\beq
\label{eq:z_to_w}
w = \l( \frac{z}{z-L} \r)^{1/n}.
\eeq
Each end-point of the subsystem $B$, $a$ and $b$, maps into $n$ points $a_k$ and $b_k$ in the complex plane, described by the coordinates
\beq\label{eq:akbk}
a_k = \l( \frac{a}{a-L} \r)^{1/n} e^{2 \pi i k/n}, \qquad  b_k = \l( \frac{b}{b-L} \r)^{1/n} e^{2 \pi i k/n}.
\eeq
The charge restricted to $B$ is 
\begin{equation}\label{eq:chargeu1}
    Q_B =\displaystyle \int_B \ dz j_0 (z)+\displaystyle \int_B \ d\bar{z} j_0 (\bar{z}),
\end{equation}
where $j_0(z)=\partial \phi(z)/(2\pi)$ and $j_0(\bar{z})=\bar{\partial} \phi(\bar{z})/(2\pi)$ have conformal dimension $(1,0)$ and $(0,1)$, respectively, and they correspond to the holomorphic ($j_0(z)$) and anti-holomorphic ($j_0(\bar{z})$) component of the bosonic current. We can focus on the holomorphic part of the charge and notice that, under the conformal map~\eqref{eq:z_to_w}, the Jacobian of the transformation simplifies and we get 
\beq
\displaystyle \int_B \ dz j_0 (z) = \displaystyle \int_B \ dw j_0 (w).
\eeq
If we exploit the fact that the theory is Gaussian, by applying the Wick theorem, we get 
\beq
\label{eq:central_Q}
\bra \exp \l( i \sum_k \gamma_k \int_{a_k}^{b_k} dw j_0(w)  \r) \ket_{\mathbb{C}}=\exp \l( - \frac{1}{2}\sum_{k,k'} \gamma_k\gamma_{k'} \int_{a_k}^{b_k} dw \int_{a_k'}^{b_k'} dw' \bra j_0(w)j_0(w')\ket_{\mathbb{C}}  \r).
\eeq

Therefore, the evaluation of the charged moments reduces to the computation of the two-point correlation function of the current field
\beq
\bra j_0 (w) j_0(w') \ket = -\frac{K}{4\pi^2(w-w')^2},
\eeq
and similarly for anti-holomorphic part $j_0({\bar{w}})$. Plugging this result in Eq.~\eqref{eq:central_Q} and taking into account both the holomorphic and anti-holomorphic parts, we get
\beq\label{eq:traces}
\dfrac{\Tr
\prod_{k=1}^{n}\Tr_{\bar{A}} [\rho e^{iQ_B \gamma_k}]}{\Tr \rho_A^n}=\exp \l( -\frac{K}{8\pi^2} \sum_{kl} \gamma_k \gamma_l M_{kl}\r),
\eeq
where $M$ is a $n \times n $ matrix whose elements are given by
\beq\label{eq:Mkl}
M_{kl} = -\int dw_1 dw_2 \frac{1}{(w_1-w_2)^2}+ \text{h.c.} = -\log \Big| \frac{(a_k-a_l)(b_k-b_l)}{(a_k-b_l)(a_l-b_k)} \Big|^2, \ k \neq l.
\eeq
Let us observe that in Eq.~\eqref{eq:traces} we have introduced the normalization factor $\Tr \rho_A^n$, which is what we expect to find if $\gamma_k\to 0$ for any value of $k$.

In order to deal with the diagonal part of the matrix $M$, we need to be more careful and properly regularize the divergence arising from $a_k\to a_l$.  
For this reason, we introduce a UV cutoff $\ep$ and regularize the distance between two coincident points as
\beq
a^{reg}_k = \l( \frac{a+\ep}{a+\ep-L} \r)^{1/n} e^{2 \pi i k/n} - \l( \frac{a-\ep}{a-\ep-L} \r)^{1/n} e^{2 \pi i k/n} \approx  \frac{- 2 \ep L e^{2 \pi i k/n}}{a^2 n  - a L n} \l( \frac{a}{a-L} \r)^{1/n},
\eeq
and similarly for $b_k^{reg}$.
Therefore, the diagonal components of the matrix $M$ read
\beq\label{eq:Mkk}
M_{kk} = -\log \Big|\frac{a^{reg}_k b^{reg}_k}{(a_k-b_k)^2} \Big|^2.
\eeq
A useful sanity check is to consider the case in which $A$ and $B$ coincide, i.e.  $[0,L]=[a,b]$. The expression for the charged moments simplifies and they read
\begin{equation}
    \Tr
\prod_{k=1}^{n}\Tr_{\bar{A}} [\rho e^{iQ_A \gamma_k}]=\Tr[\rho_Ae^{iQ_A \sum_k\gamma_k}]
\end{equation}
This object has played an important role in the computation of the symmetry resolution of the entanglement~\cite{goldstein}, and their analytical expression is well known for a compact boson. By taking $a=\ep, b=L+\ep$, one can check that the $L$-dependent part reads $M_{kl}=4/n\log \frac{L}{\ep}$ for $k,l=1,\dots,n$ and 
\begin{equation}
    \exp \l( -\frac{K}{8\pi^2} \sum_{kl} \gamma_k \gamma_l M_{kl}\r)=\exp \l( -\frac{K}{2\pi^2}\log \frac{L}{\ep} (\sum_{k} \gamma_k)^2\r).
\end{equation}
We remark that here we are neglecting an $L$-independent contribution which makes our result different with respect to the charged moments associated with the symmetry-resolved entanglement. The reason comes from the regularization that we are taking in terms of the UV cutoff $\epsilon$, such that the current operator implementing the charge does not lie exactly at the entangling point but it is a bit deviated from it.

Once we plug the result~\eqref{eq:traces} into Eq.~\eqref{eq:tr2}, we find that 
\begin{equation}
\label{eq:final}
    \int d\gamma_1\cdots d\gamma_n \exp \l(- \frac{K}{8\pi^2} \sum_{kl} \gamma_k \gamma_l M_{kl}-i\sum_j\gamma_j q\r)\simeq \frac{e^{-2\pi^2 q^2 C_n/K}}{\sqrt{{\rm det} M}} \l( \frac{8 \pi^3}{K} \r)^{n/2},
\end{equation}
where $C_n=v M^{-1} v^T$, $v$ is a $n$-dimensional vector $(1 \dots 1)$ and, to solve the integral, we have used the saddle point approximation. This is justified by the structure of the matrix $M$ that we are going to study in great detail in the next paragraph.

\paragraph{Analytical details about the matrix $M$:}
We can write explicitly the matrix elements of $M$ in terms of the parameters $a,b$ and $L$ as
\begin{equation}\label{eq:Mjk}
    M_{jk}=\begin{cases} \tilde{a}_{j-k}\equiv -2 \log \left[\frac{4 \left(\frac{a}{a-L}\right)^{\frac{1}{n}} \left(\frac{b}{b-L}\right)^{\frac{1}{n}} \sin ^2\left(\frac{\pi 
   (j-k)}{n}\right)}{-2 \left(\frac{a}{a-L}\right)^{\frac{1}{n}} \left(\frac{b}{b-L}\right)^{\frac{1}{n}} \cos \left(\frac{2 \pi 
   (j-k)}{n}\right)+\left(\frac{a}{a-L}\right)^{2/n}+\left(\frac{b}{b-L}\right)^{2/n}}\right],\quad &j\neq k\\
  \tilde{a}_0\equiv -2 \log \left[\frac{4 L^2\epsilon^2\left(\frac{a}{a-L}\right)^{\frac{1}{n}} \left(\frac{b}{b-L}\right)^{\frac{1}{n}} }{a b n^2 (a-L)(b-L)[(\frac{a}{a-L})^{1/n}-(\frac{b}{b-L})^{1/n}]^2}\right], \quad &j= k
    \end{cases}
\end{equation}
This equation shows that $M$ is a symmetric circulant matrix~\cite{circulant}, and we can use this property in order to evaluate both the inverse matrix $M^{-1}$ and the determinant. First of all, the elements of the inverse matrix are given by 
\begin{equation}
    M^{-1}_{jl}=\frac{1}{n}\sum_{k=0}^{n-1}\frac{e^{2\pi i (j-l)k/n}}{\sum_m \tilde{a}_me^{2\pi i k m/n}},
\end{equation}
and it is also useful to compute 
\begin{equation}
   \sum_jM^{-1}_{jl}=\frac{1}{n}\sum_j\sum_{k=0}^{n-1}\frac{e^{2\pi i (j-l)k/n}}{\sum_m \tilde{a}_me^{2\pi i k m/n}}=\frac{1}{\sum_m \tilde{a}_m}. 
\end{equation}
Putting everything together, we get
\begin{equation}\label{eq:Cn}
    C_n=n\left(\sum_j \tilde{a}_j\right)^{-1}=\frac{n}{4\log[(b-a)/(2\epsilon)]}.
\end{equation}
Surprisingly, the dependence on the replica index considerably simplifies and $C_n$ is $L$-independent.
We can also write down an analytical expression for the determinant in terms of the matrix elements
\begin{equation}   \label{eq:detM}\mathrm{det}M=\prod_{k=0}^{n-1}\left[\tilde{a}_0+\sum_{j=1}^{n-1}\tilde{a}_j
e^{2\pi i jk/n}\right].\end{equation}
A closer inspection of the matrix elements in Eq.~\eqref{eq:Mjk} shows that $\tilde{a}_0\gg \tilde{a}_j$, and, as a consequence, $\mathrm{det}M\approx \tilde{a}_0^n$.
In order to completely compute Eq.~\eqref{eq:tr2}, we also need to take into account the normalization factor $\mathcal{N}$. %since so far we only computed the ratio $  \Tr\prod_{k=1}^{n}\Tr_{\bar{A}} [\rho e^{iQ_A \gamma_k}]/   \Tr\rho_A^n$, we need to normalize to make $\Tr \rho_A=1$. 
From the results above, we find 
\begin{equation}   \mathcal{N}=\frac{e^{-2\pi^2 q^2 C_1/K}}{\sqrt{M_{11}}},
\end{equation}
and, therefore,
\beq\label{eq:step4}
\frac{\Tr \rho_{A,q}^n}{\mathrm{Tr}\rho_A^n} = \frac{e^{-2\pi^2 q^2 C_n/K}}{\sqrt{\det M}}  \left(  \frac{e^{-2\pi^2 q^2 C_1/K}}{\sqrt{M_{11}}} \right)^{-n}.
\eeq
The denominator $\mathrm{Tr}\rho_A^n$ is the usual partition function on the $n-$sheeted Riemann surface which, for this geometry, reads $\mathrm{Tr}\rho_A^n\propto L^{-\frac{1}{6} (n - \frac{1}{n})}$.
Given the simple dependence on $n$ of Eq.~\eqref{eq:Cn}, the formula~\eqref{eq:step4} gives
\begin{equation}\label{eq:sumq}
   \frac{\Tr \rho_{A,q}^n}{\mathrm{Tr}\rho_A^n}=\sqrt{\dfrac{M_{11}^n}{\mathrm{det}M}}, \quad {\rm MIE}(A: \bar{A\cup B})=\sum_q p_q S^{(n)}_A( \Pi^B_q | \Psi \ket)=S_A^{(n)}+\dfrac{1}{2(1-n)}\log\dfrac{M_{11}^n}{\mathrm{det}M}.
\end{equation}
Using the structure of the determinant of $M$, the subleading term in the result above is negative and very close to 0, since $\det M\approx a_0^n=M_{11}^n$. The minus sign reflects that the measurement reduces our system’s uncertainty, and therefore its entropy. 

Let us now analyze the physical consequences of this result. The effect of projecting into a given charge sector in part of a system almost disappears when we sum over all of them. Therefore, at leading order, we simply retrieve the total entanglement entropy between $A$ and $\bar{A}$, up to a small correction depending on the geometry of the measured region. This is somehow expected because the operation that we are doing on our system is not as drastic as performing projective measurements, and the total entanglement structure is only mildly affected by this.

\begin{figure}[ht!]
\centering
    {\includegraphics[width=0.323\textwidth]{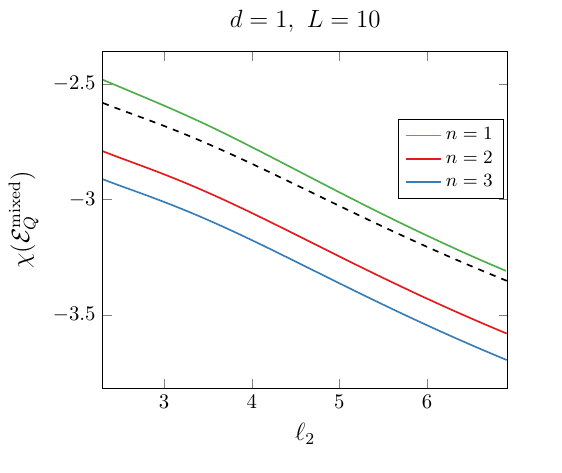}}
    {\includegraphics[width=0.323\textwidth]{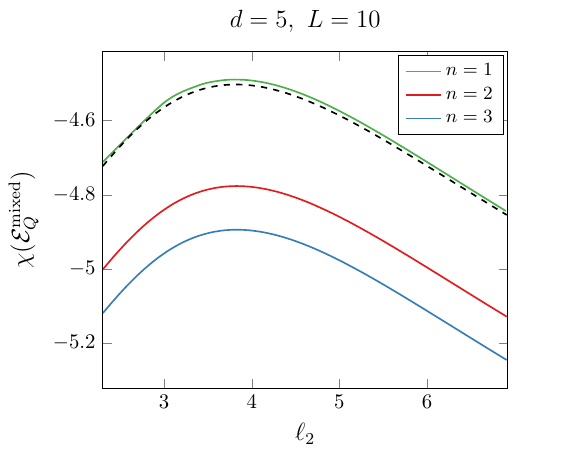}}
    {\includegraphics[width=0.323\textwidth]{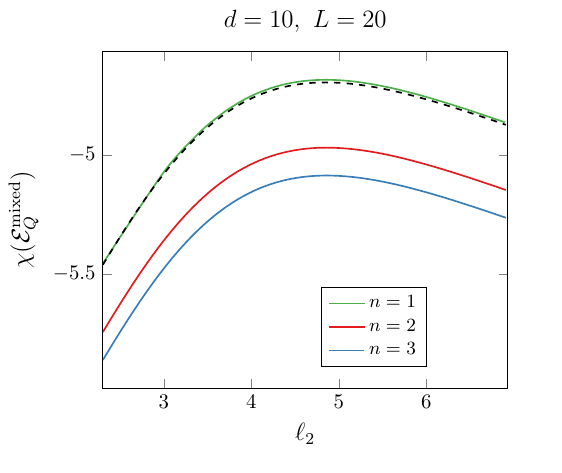}}
     \caption{Log-log plot of the Holevo $\chi$-quantity as a function of $b-a=\ell_2$ for different values of subsystem size of $A$, $L$, and different $a=d+L$, i.e. different distances between $A$ and $B$. The $\chi$-quantity corresponds to the green line $n=1$ and it has been obtained by using a numerical approach. The black dashed line is the approximation in Eq.~\eqref{eq:approx}, valid for $\ell_2\gg \epsilon$.
     As a reference, we also plot different values of $n$ for $\frac{1}{2(1-n)}\log\frac{M_{11}^n}{\mathrm{det}M}$.}
     \label{fig:holevo}
\end{figure}

To have a more refined analysis about the bound on the accessible information we can extract from our charge-projected ensemble, we can evaluate the Holevo $\chi$-quantity, which reads from Eq.~\eqref{eq:sumq}
\begin{equation}\label{eq:chi1}
   \chi(\mathcal{E}^{\rm mixed}_\mathcal{Q} ) \equiv S_A^{(1)} (\rho_{A} ) - \sum_q p_q S^{(1)}_A( \Pi^B_q | \Psi \ket)=\lim_{n\to 1} \dfrac{1}{2(1-n)}\log\dfrac{M_{11}^n}{\mathrm{det}M}.
\end{equation}
We remark that, since we are simply subtracting the entanglement entropy of the subsystem $A$, $S_A^{(1)} (\rho_{A} )$--which does not depend on the measurements--studying the change induced in our ensemble from Eq.~\eqref{eq:sumq} or Eq.~\eqref{eq:chi1} is completely equivalent.
However, studying Eq.~\eqref{eq:chi1} would require performing an analytic continuation that we do not know how to evaluate. 
We can bypass this problem by using a numerical approach~\cite{aaa}. We report the result of this numerical analysis in Fig.~\ref{fig:holevo} by fixing the size of $A$, $L$, the position $a=d+L$ and varying $\ell_2$, which denotes the size of the measured region. We observe that as far as $A$ and $B$ are distant, the Holevo-quantity increases for small values of $\ell_2$, and then it slowly decreases. Let us fix the size of $A$ and the distance between $A$ and the left-most point of $B$, $d$. If $B$ is very small then performing the measurement on it does not alter the original state much, so it is hard to guess anything just by studying the post-measurement state in $A$. At the same time, if $B$ is very large then the information about the measurement inside $B$ will be weakly correlated with $A$ since there are parts of $B$ which are very far away from $A$. In other words, there exists an optimal size of $B$ from which we can `maximally' extract information and reconstruct the measurement outcome $q$, but after this threshold, since the subsystem size of $A$ is fixed, the more we measure $B$, the less 
information we gain. Notice that in the opposite regime, if we keep $B$ fixed but increase $A$, the amount of information grows monotonically. This can be extracted from our analytical approximation (\ref{eq:approx}) below. 
%If $\ell_2$ is small enough and $A$ and $B$ are distant, we can explain the increase in the upper bound of the accessible information by noting that, in the system we are considering, the correlations between $A$ and $B$ decay algebraically, so it means that the farther they are, the larger is the size of $B$ we need to measure to gain information about $q$ (having access only to $\rho_{A,q}$), at least before the threshold above mentioned.}

We also remark here an important difference between our setup and the case of a standard projected ensemble: after our measurement and partial trace over $\bar{A}$, we get an ensemble of mixed states and the bound on the accessible information depends on the size of the subsystem $B$. On the other hand, for the projected ensembles described by Eq.~\eqref{eq:proj_ens}, the Holevo $\chi$-quantity reduces to the standard entanglement entropy of $\rho_A$, so it does not depend at all on the size of the projected region.

If we focus on the regime $\tilde{a}_0\gg \tilde{a}_j$ in Eq.~\eqref{eq:Mjk}, we can obtain a close expression for the analytical continuation above. Indeed, the second term in Eq.~\eqref{eq:sumq} can be rewritten as
\begin{equation}
\begin{split}
    \frac{\log \det M-n\log M_{11}}{2(n-1)}&=-\frac{1}{2(n-1)}\sum_{k=0}^{n-1}\log [1+\sum_{j=1}^{n-1}e^{2\pi i kj/n}\frac{\tilde{a}_j(n)}{\tilde{a}_0(n)}]+\frac{n}{2(n-1)}\log \frac{\tilde{a}_0(1)}{\tilde{a}_0(n)}\\
    &\stackrel{\tilde{a}_0\gg \tilde{a}_j}{\simeq} -\frac{1}{2(n-1)}\sum_{k=0}^{n-1}\sum_{j=1}^{n-1}e^{2\pi i kj/n}\frac{\tilde{a}_j(n)}{\tilde{a}_0(n)}+\frac{n}{2(n-1)}\log \frac{\tilde{a}_0(1)}{\tilde{a}_0(n)}.
    \end{split}
\end{equation}
The only non-vanishing contribution of the expression above comes from the second term because the sum over $k$ imposes $j=0$, which is absent, such that 
\begin{equation}\label{eq:approx}
    \lim_{n\to 1} \frac{1}{2(n-1)}\log \frac{\det M}{M^n_{11}}\simeq \frac{1}{\log
   \left(\frac{b-a}{2\epsilon}\right)}-\frac{(2 a b-L (a+b)) \log \left[\frac{a (b-L)}{b (a-L)}\right]}{2 L (b-a) \log
   \left(\frac{b-a}{2\epsilon}\right)}.
\end{equation}
We report this approximation for the analytical continuation as black dashed lines in Fig.~\ref{fig:holevo}. We observe that, as we increase the distance between $A$ and $B$ and $\ell_2\gg \epsilon$, the approximation get closer and closer to the exact solution.

We can slightly generalize the result above by considering a dynamical setup, where we measure the charge of the subsystem $B$ at time $t-i\epsilon'$. This means that we have to be careful when we consider the anti-holomorphic part. Indeed, the chiral components can be obtained from Eq.~\eqref{eq:akbk}, by shifting $a,b$ as $a\to a+i(\tau-\epsilon')$ and $b\to b+i(\tau-\epsilon')$, while for the anti-holomorphic components we should consider $a\to a+i(\tau+\epsilon')$ and $b\to b+i(\tau+\epsilon')$. Therefore, the matrix $M$ is $2n\times 2n$, it acquires a non-trivial time dependence but preserves its symmetric circulant structure. In the large time limit, we can compute
\begin{equation}
    \lim_{n\to 1} \frac{1}{2(n-1)}\log \frac{\det M(t)}{M^n_{11}(t)}\stackrel{t\to \infty}\simeq \frac{(b-a)^2 L^2}{24\log[(b-a)/(2\epsilon)]t^4},
\end{equation}
which implies that the measurement-induced entanglement in Eq.~\eqref{eq:sumq} simply reduces to the total von Neumann entropy between $A$ and $\bar{A}$, $S_A^{1}$, and the accessible information goes to 0. In other words, at large time, the effect of the partial measurement of the charge vanishes, as one would have expected. 

\section{Projecting other operators}
\label{sec:others}

In this section, we will consider the situation where we project some other operator, not necessarily the charge. However, we will still assume that it is extensive
\beq
Q_B = \int_B dx \ \Oc(x),
\eeq
and
$\Oc$ is a Hermitian operator. It can be either a scalar primary $\Oc_{s}(z,\bar{z})$ of dimension $(h_s/2,h_s/2)$ or a non-conserving vector $j_0(z,\bar{z})$, which is a sum of two operators of weight $(1+h_v/2,h_v/2)$ and $(h_v/2,1+h_v/2)$ to make it Hermitian.   The real $x$ variable we are using parameterizes the $t=0$ slice, $z=\bar{z}=x$.
Finally, we should keep in mind that $h_{s,v}$ cannot be too large, since irrelevant operators in QFT are very UV-sensitive. We will make this statement more precise below.

The steps we need to perform to study the observable-projected ensemble are the same as in the previous section.
The reduced density matrix $\rho_{A,q}$ after the projection is defined in exactly the same way (c.f. Eq.~\eqref{eq:rhoAq_simp})
\beq
\rho_{A,q} = \frac{1}{\Nc}  \int d\gamma  \ e^{-i \gamma q} \Tr_{\bar{A}}( e^{i \gamma Q_B}\rho ).
\eeq
In evaluating the Renyi entropies $\Tr \rho_{A,q}^n$, we can again perform a conformal transformation $w(z)$, as in Fig.~\ref{fig:cft_sheets}, to map it to a correlation function on a sphere.
However, now we have to take into account the Jacobian factors. For example, for a primary operator of dimension $(h_s/2,h_s/2)$, we need to evaluate 
\beq
\bra \exp \l( i \sum^n_{k=1} \gamma_k \int_a^b dx \ \l( \frac{\pr w_k}{\pr x} \frac{\pr \bar{w}_k}{\pr x}\r)^{h_s/2} \Oc_{s}(w_k(x),\bar{w_k}(x)) \r)  \ket_{\mathbb{C}},
\eeq
where for convenience we kept the original integration variable $x$ and $w_k$ are different branches of the conformal mapping $w(z)$
\beq\label{eq:map_gen}
w_k = e^{2 \pi i k/n} \l( \frac{z}{z-L} \r)^{1/n}.
\eeq

 Without additional assumptions, it would be very hard to evaluate these expectation values.
The only general statement we can make is that if $a \ra b$ or $L \ra 0$ we get the operator product expansion (OPE) limit in which the correlations of the measured observables among different replicas decouple and we simply obtain the standard R\'enyi entropies.
 
 We can derive significant results if we assume that $\Oc$ has Gaussian correlators. This assumption is not very restrictive since several models satisfy this criterion, such as free theories, holographic theories, and the so-called symmetric orbifold CFTs. The latter are defined as follows \cite{Dijkgraaf:1996xw,Lunin:2000yv}: we can take $N$ copies of any ``seed" CFT $T$, which we denote as $T^{\otimes n}$ and consider the following quotient:
\beq
\text{Sym orbifold}(T) = \frac{T^{\otimes N}}{\text{Sym}_N}.
\eeq
An example of an operator in this theory (sometimes called untwisted sector) is
\beq
\Oc_s = \frac{1}{N} \sum_{i=1}^N \Oc_i,
\eeq
where $\Oc_i$ belong to different copies. Thanks to the central limit theorem the correlations of $\Oc_s$ are Gaussian for large $N$.

If $\Oc$ are Gaussian operators, then we can again use the Wick contractions to arrive at 
\beq\label{eq:gaussian_gen}
\dfrac{\Tr
\prod_{k=1}^{n}\Tr_{\bar{A}} [\rho e^{iQ_B \gamma_k}]}{\Tr \rho_A^n}=\exp \l( -\frac{1}{2} \sum_{ij} \gamma_i \gamma_j M_{ij}
\r),
\eeq
where $M$ depends on the type of operator we are studying.

Beyond the Gaussian fields, we can look at Eq.~\eqref{eq:gaussian_gen} as a leading perturbative expansion in $\gamma$. In section~\ref{sec:num_checks}, we will check our predictions against the numerical computations in the Majorana CFT where the charge field is indeed non-Gaussian. Also, we will see in section~\ref{sec:universal} that such quantities with fixed $\gamma$ (charge distribution generating functions) have better UV properties.

For $\Oc= \Oc_{s}(z,\bar{z})$, the matrix elements are given by
\beq\label{eq:scalar}
M_{ij} = \int dx_1 dx_2 \bra \Oc_{s}(w_i(x_1),\bar{w_i}(x_1)) \Oc_{s}(w_j(x_2),\bar{w_j}(x_2))  \ket \l( \frac{\pr w_i}{\pr x_1} \frac{\pr \bar{w_i}}{\pr x_1}  \r)^{h_s/2}
\l( \frac{\pr w_j}{\pr x_2} \frac{\pr \bar{w_j}}{\pr x_2}  \r)^{h_s/2},
\eeq
while for the vector $j_0$:
\beq\label{eq:vector}
\begin{aligned}
M_{ij} = \int dx_1 dx_2 \bra j_0(w_i(x_1),\bar{w_i}(x_1)) j_0(w_j(x_2),\bar{w_j}(x_2))  \ket \times \nonumber \\
\times \l( \frac{\pr w_i}{\pr x_1} \r)^{1+h_v/2} \l( \frac{\pr \bar{w_i}}{\pr x_1} \r)^{h_v/2} \l( \frac{\pr w_j}{\pr x_2} \r)^{1+h_v/2} \l( \frac{\pr \bar{w_j}}{\pr x_2} \r)^{h_v/2}
+h.c.
\end{aligned}
\eeq

In the $w$ plane, the correlation function of $\Oc$ can be easily evaluated in the ground state 
\beq
\bra \Oc_{s}(w_i,\bar{w_i}) \Oc_{s}(w_j,\bar{w_j}) \ket = 
\frac{1}{|
w_i-w_j|^{2h_s}
},
\eeq
\beq
\bra j_0(w_i,\bar{w_i}) j_0 (w_j,\bar{w_j}) \ket = 
-\frac{1}{(w_i-w_j)^2
|w_i-w_j|^{2h_v}  
}.
\eeq
For simplicity, we have normalized them to $1$, since a global prefactor would only affect the overall  ``charge" distribution variance. The signs are fixed by requiring reflection-positivity in the Euclidean spacetime. Once we plug these expressions in the matrix elements $M_{ij}$, the resulting integrals can only be evaluated numerically.
However, we can still discuss a few general features of these expressions.

The integrals~\eqref{eq:scalar},~\eqref{eq:vector} are finite for $i \neq j$ because $w_i\neq w_j$, while for $i=j$, the correlators diverge and the resulting singularity might not be integrable. However, this divergence is universal in the following sense: since it comes from two points colliding, the divergence structure does not depend on the number of replicas or the interval lengths. 
For example, for the $(h_s/2,h_s/2)$ case the integrand for $M_{ii}$ has the following form at $x_1 \approx x_2$:
\beq
\begin{aligned}
\bra
\Oc_{s}(w_j(x_1),\bar{w_j}(x_1)) \Oc_{s}(w_j(x_2),\bar{w_j}(x_2))  \ket \l( \frac{\pr w_j}{\pr x_1} \frac{\pr \bar{w_j}}{\pr x_1}  \r)^{h_s/2}
\l( \frac{\pr w_j}{\pr x_2} \frac{\pr \bar{w_j}}{\pr x_2}  \r)^{h_s/2} \approx \nonumber \\ \approx 
\frac{1}{(x_1-x_2)^{2h_s}} \l( 1 + \frac{h_s L (n^2-1) (x_1-x_2)^2}{12n^2 (L-x_2)^2 x_2^2} + \dots \r).
\end{aligned}
\eeq
Hence, if the operator is not too irrelevant, that is $h_s<3/2$, we can remove the divergence by subtracting $1/(x_1-x_2)^{2h_s}$, which is simply the two-point function of $\Oc_{s}(x)$ on a plane. Correspondingly, the integral of $1/(x_1-x_2)^{2h_s}$ controls the distribution of $Q_B$ \textit{for a single interval}. The same conclusion holds for the vector case if we impose $h_v<1/2$. We will return to this observation later when we discuss how to define UV-finite quantities, while now we evaluate the entanglement entropy. We also notice that, if $h_s<1/2$, then UV divergences completely disappear.

Similar to the charge case, we can integrate over $\gamma_k$ with the weight $\exp(-i q \sum_k \gamma_k)$ and we get
\beq\label{eq:gauss_gen2}
\frac{\Tr \rho_{A,q}^n}{\mathrm{Tr}\rho_A^n} = \frac{e^{-q^2 C_n/2}}{\sqrt{\det M}}  \left(  \frac{e^{-q^2 C_1/2}}{\sqrt{M_{11}}} \right)^{-n},
\eeq
where $C_n = (1,\dots)^T M^{-1} (1,\dots)$.
Subtracting the UV-divergent terms at this stage would be very difficult because $C_n$ is highly non-linear in $M$.

\begin{figure}
{\includegraphics[width=0.495\textwidth]{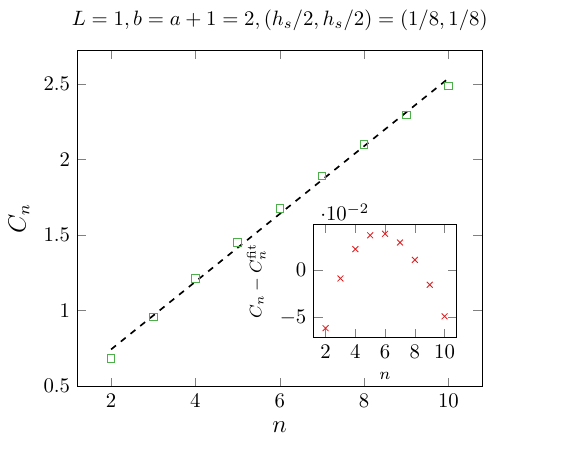}}  
{\includegraphics[width=0.495\textwidth]{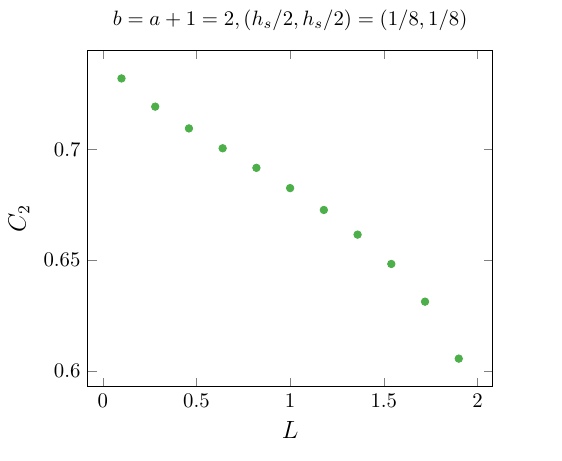} } 
    \caption{Plot of the $C_n$ coefficient defined in Eq.~\eqref{eq:gauss_gen2}. In the left panel, we prove that its dependence on $n$ is nonlinear, as the disagreement between the exact numerical values of $C_n$ (symbols) and the best linear fit (dashed black line) shows. The inset also reveals a discrepancy $O(10^{-2})$ between the numerical data and the linear fit. In the right panel, we plot the non-trivial $L$ dependence of $C_2$ for the 2-replica computation.
    In both cases, we have used the operator with conformal weight $(1/8,1/8)$. }
    \label{fig:nontriv}
\end{figure}

Unlike the case in which the observable is simply the charge operator of a compact boson, (see Eq.~\eqref{eq:Cn}), we could not find an explicit formula for $C_n$. However, it is relatively easy to find it numerically. Two important features, which are not present in the charge case~\eqref{eq:Cn}, are: 
\begin{itemize}
    \item $C_n$ is not linear in $n$ - Fig.~\ref{fig:nontriv} (left panel). 
    \item $C_n$ depends on the length $L$ of the interval $A$ (the one for which we evaluate the entanglement entropy)
    - Fig.~\ref{fig:nontriv} (right panel).
\end{itemize}
The non-linearity of $C_n$ implies that the $q$-dependence of Eq.~\eqref{eq:gauss_gen2} does not trivially simplify, as it happened for the compact boson. Moreover, we have numerically checked that $C_n$ is not even a function of the conformal cross-ratio of the two intervals $A$ and $B$.

If we sum over all the possible measurement outcomes $q$, we get
\begin{equation}\label{eq:sumqgen}
 {\rm MIE}(A: \bar{A\cup B})\simeq S_A^{(n)}+\dfrac{1}{2(1-n)}\log\dfrac{M_{11}^n}{\mathrm{det}M}-\frac{C_n-nC_1}{2\sqrt{2\pi C_1^3 M_{11}}(1-n)},
\end{equation}
where we have used that $\sum_q p_q q^2\simeq (2\pi C_1^3 M_{11})^{-1/2}$.
Contrarily to the result~\eqref{eq:sumq}, the equation above displays a non-trivial dependence on the size of $A$ and on the measured region through the last term.

It is important to discuss the relevant point-splitting procedure for the evaluation of the integrals~\eqref{eq:scalar},~\eqref{eq:vector}. The main point is that it has to preserve the current conservation (e.g. $\pr_{\bar{z}} j_0(z)=0$), which means that the regularization must not mix holomorphic and anti-holomorphic parts.
For example, the integrand for the two-point function
of a conserved current appearing in Eq. \eqref{eq:Mkl} must be regularized as
\beq\label{eq:div1}
\int dz_1 dz_2 \ \frac{1}{(z_1-z_2)^2}  \ra 
\int dz_1 dz_2 \ \frac{1}{(z_1-z_2+ i \ep)^2}.
\eeq
%which produced Eq. (\ref{eq:Mkl}). 
Alternatively, one can understand it as adding an extra $i \epsilon$ Euclidean time evolution before the operators are inserted.

Now, for the scalar operator of dimension $(h_s/2,h_s/2)$ the integral after regularization becomes
\beq
\label{eq:conf_int}
\int_{a}^b \frac{dz_1 dz_2}{|z_1 - z_2|^{2h_s}} \ra 
\int_{a}^b \frac{dz_1 dz_2}{(|z_1 - z_2|^{2}+\ep^2)^{h_s}} =\frac{1}{-1+2h_s}
|b-a| \ep^{1-2h_s} + 
\frac{|b-a|^{2-2h_s}}{1 - 3 h_s + 2 h_s^2}+\Oc(\ep).
\eeq
Some special cases are $h_s=1/2$
\beq
\int_{a}^b \frac{dz_1 dz_2}{|z_1 - z_2+ i \ep|} =2(b-a) \l( \log \frac{|b-a|}{\ep} - 1 \r), 
\eeq
and $h_s=1$
\beq
\label{eq:ising}
\int_{a}^b \frac{dz_1 dz_2}{|z_1 - z_2+ i \ep|^2} = \frac{\pi |b-a|}{\ep} - 2 \l( 1+\log \frac{|b-a|}{\ep} \r) + \Oc(\ep).
\eeq
For the vector case, we have the following regularized integral:
\beq
\int_{a}^b dz_1 dz_2 \frac{-1}{(z_1-z_2+ i \ep)^2 (|z_1 - z_2|^{2}+\ep^2)^{h_v }},  
\eeq
which is much harder to evaluate. 
For small $\ep$, it yields a series of terms starting with $\ep^{-1-2h_v}$ with non-universal coefficients plus a regular term $|b-a|^{-2h_v}$ with a universal coefficient. One can adopt a slightly different regularization where this universal term can be easily evaluated for $h_v> 0$:
\beq
\int_{a}^b dz_1 dz_2 \frac{-1}{(z_1-z_2+ i \ep)^2 |z_1 - z_2|^{2h_v }}  = -\frac{\const}{\ep^{2h_v}} + \dots + \frac{|b-a|^{2h_v}}{2h_v(1+2h_v)} + \Oc(\ep).
\eeq
\section{Extracting universal quantities}
\label{sec:universal}
In the previous sections, we showed that computing the entropy after the projections produces UV divergences if we measure an operator of large conformal dimensions. These divergences, however, are tied to colliding operators and are replica-diagonal. Thus it seems possible to subtract them in a physical way. In this section, we discuss how to do it by studying quantities related to an observable-projected ensemble.

The probabilities $p_q$ of observing different charges, or more generally extensive quantities $\int_B dx \ \Oc$ inside the region $B$, are determined by the Born rule
\beq
p_q = \Tr (\Pi_q \rho).
\eeq
Summing over $q$ with $e^{i \gamma q}$ weight produces the generating function $\bra e^{i \gamma Q_B} \ket$,
\beq
\sum_q p_q e^{i \gamma q} = \Tr( \rho e^{i \gamma Q_B} ).
\eeq

By preparing two copies of the system and measuring $q_1, q_2$ inside the two copies of $B$, we can access the overlap (for example, by performing a swap test)
\beq\label{eq:distance}
R_{q_1 q_2} = \Tr(\rho_{A,q_1} \rho_{A,q_2}),
\eeq
where the post-measurement state of $A$ is
\beq
\rho_{A,q} = \frac{1}{p_q} \Tr_{\bar{A}} (\Pi_q \rho \Pi_q) \equiv \frac{1}{p_q} \tilde{\rho}_{A,q},
\eeq
with $\tilde{\rho}_{A,q}$ the unnormalized density matrix
\beq\label{eq:rhotilde}
\tilde{\rho}_{A,q} = \frac{1}{(2 \pi)^2}\Tr_{\bar{A}} \int d\gamma_1 d\gamma_2 \ e^{i \gamma_1 q - i \gamma_2 q} e^{-i \gamma_1 Q_B} \rho e^{+ i \gamma_2 Q_B}.
\eeq
We remark that Eq.~\eqref{eq:distance} represents the overlap between two operators, $\rho_{A,q_1}$ and  $\rho_{A,q_2}$, so measuring $R_{q_1 q_2}$ can be relevant to understand how much two operators are distinguishable if $q_1\neq q_2$.
Since $B \in \bar{A}$, we can move $e^{i \gamma_i Q_B}$ inside the trace in Eq.~\eqref{eq:rhotilde} to obtain
\beq
\tilde{\rho}_{A,q} = \frac{1}{2\pi} \Tr_{\bar{A}} \int d\gamma \  e^{-i \gamma q} \rho e^{i \gamma Q_B} \equiv \frac{1}{2\pi}  \int d\gamma \ e^{-i \gamma q} \tilde{\rho}_{A,\gamma}.
\eeq

Therefore, if we prepare the system samples and compute the average $R_{q_1 q_2}$ over the charges with weights $e^{i \gamma_1 q_1 + i \gamma_2 q_2}$, it will yield
\begin{align}\label{eq:rel}
\sum_{q_1 q_2} e^{i \gamma_1 q_1 + i \gamma_2 q_2} p_{q_1} p_{q_2} R_{q_1 q_2} = \sum_{q_1 q_2} e^{i \gamma_1 q_1 + i \gamma_2 q_2} \Tr_A \l( \tilde{\rho}_{A,q_1} \tilde{\rho}_{A,q_2} \r) = \\
=\frac{1}{(2\pi)^2}  \sum_{q_1 q_2 }e^{i \gamma_1 q_1 + i \gamma_2 q_2}
\int d\gamma_1 d\gamma_2 e^{-i q_1 \gamma_1 - i q_2 \gamma_2} \Tr_A ( \tilde{\rho}_{A,\gamma_1} \tilde{\rho}_{A,\gamma_2}).
\end{align}
By explicitly summing over $q_1, q_2$, Eq.~\eqref{eq:rel} gives
\beq\label{eq:rel1}
\sum_{q_1 q_2} e^{i \gamma_1 q_1 + i \gamma_2 q_2} p_{q_1} p_{q_2} R_{q_1 q_2} =\Tr (\tilde{\rho}_{A,\gamma_1} \tilde{\rho}_{A,\gamma_2}),
\eeq
which is a two-replica calculation with the appropriate fluxes. As we discussed before, the UV divergences arise only within each replica copy and they are the same as in computing $\bra e^{i \gamma_1 Q_B}\ket$. Even though, strictly speaking, our arguments are based on an explicit Gaussian computation, in Sec.~\ref{sub:interac}, we will argue that they hold more generally. Hence, 
if we divide Eq.~\eqref{eq:rel1} by $\bra e^{i \gamma_1 Q_B} \ket \bra e^{i \gamma_2 Q_B} \ket$, the UV-divergence will cancel and
\beq
\frac{\Tr (\tilde{\rho}_{A,\gamma_1} \tilde{\rho}_{A,\gamma_2})}{\bra e^{i \gamma_1 Q_B} \ket \bra e^{i \gamma_2 Q_B} \ket}  \text{  is UV-finite}.
\eeq

The above observable is unusual because it includes the overlap of two different states and it might be difficult to access it theoretically or experimentally. 

A more standard object to study is the averaged purity:
\beq
\sum_q p_q e^{i\gamma q} \Tr(\rho_{A,q}^2) =  \sum_q \int d\gamma_1 d\gamma_2 
e^{i \gamma q - i \gamma_1 q - i \gamma_2 q} \Tr(\rho_{A,\gamma_1} \rho_{A,\gamma_2}).
\eeq
Let us repeat the analysis above to see if it can be made finite by dividing by a correlation function involving $e^{i Q_B}$.
Now the sum over $q$ projects $\gamma_1 + \gamma_2 =\gamma$, so we need to compute the integral
\beq
\int d\gamma_1 d\gamma_2 \ \delta(\gamma-\gamma_1-\gamma_2) e^{-M_{11} \gamma_1^2/2 -  \gamma_1 \gamma_2 M_{12} - M_{22} \gamma_2^2/2}.
\eeq
After simple manipulations, it leads to
\beq\label{eq:rel2}
\sum_q p_q e^{i\gamma q} \Tr(\rho_{A,q}^2) =\frac{\sqrt{\pi}}{\sqrt{M_{11}-M_{12}}} \exp \l( -\frac{\gamma^2}{4} (M_{11}-M_{12}) \r).
\eeq
Hence we can still find a UV-finite quantity if we divide Eq.~\eqref{eq:rel2} by $\bra e^{i \gamma  Q_B/\sqrt{2}} \ket$.

\subsection{Introducing interactions}\label{sub:interac}

\begin{figure}
    \centering
    \includegraphics[scale=1.5]{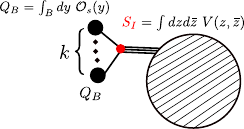}
\caption{A generic Feynman diagram contributing to $\bra Q_B^k \ket$ . Black dots indicate the insertion of $Q_B$ and single black lines are $\Oc_s$ propagators. The red dot is the insertion of the interaction $S_I$. Multi-line emanating from it represent all possible fields leaving that vertex. The striped disk contain the rest of the diagram.}
\label{fig:generic_diagram}
\end{figure}

In the previous part, we explained that the ratio
\beq
\label{eq:ratio2}
\frac{\Tr (\tilde{\rho}_{A,\gamma_1} \tilde{\rho}_{A,\gamma_2})}{\bra e^{i \gamma_1 Q_B} \ket \bra e^{i \gamma_2 Q_B} \ket} 
\eeq
is UV-finite and our arguments were based on the explicit expression for Gaussian theories found in Sec.~\ref{sec:others}. Now, we would like to argue that it is true more generally.
The idea is that interactions introduce extra smearing, so the diagrams are less divergent. 
In other words, diagrams that involve interactions are finite.
We can illustrate this by evaluating the expectation value $
\bra e^{i \gamma Q_B} \ket$ by expanding it in the powers of $\gamma$.
 Let us imagine that we are drawing all possible Feynman diagrams associated with this quantity. As usual, disconnected diagrams combine into an exponent of connected ones \cite{Peskin:1995ev}, so we only need to focus on connected diagrams. A generic connected diagram is illustrated in Fig.~\ref{fig:generic_diagram}. Importantly, we will assume that the interaction vertex $V(z,\bar{z})$ can be connected to $k$ copies of $Q$ and that it has no extra derivatives, so the external propagators are simply the $\Oc_s$ Green functions. The analysis below can be easily modified to include these possibilities.
As in the rest of the paper we can consider two cases: measuring a scalar operator of the conformal weight $(h_s/2,h_s/2)$ or a vector with weights $(1+h_v/2 ,h_v/2)$ and $(h_v/2,1+h_v/2)$. 

We can start from the scalar case, where
the external propagator is 
\beq\label{eq:scalar_int}
\int_{B=[a,b]} dy \ \frac{1}{((y-z) (y-\bar{z}))^{h_s}},
\eeq
and $z,\bar{z}$ are the coordinates of the interaction vertex represented by the red blob. If we assume that $z$ is outside the interval $B$, the integral over $y$ can be computed exactly and it gives
\beq
\label{eq:ext_int_hh}
-(-1)^{h_s} \frac{1}{1-h_s}(y-\bar{z})^{1-h_s} \l( \frac{1}{z-\bar{z}}\r)^{h_s}
\phantom{m}_2 F_1 \l( 1-h_s,h_s,2-h_s,\frac{y-\bar{z}}{z-\bar{z}} \r) \Big |^{y=b}_{y=a}.
\eeq

Since there might be several $Q_B$ insertions leading to the same vertex, Eq. \eqref{eq:ext_int_hh} must be taken to some power $k$. The second step is to perform the integral over $z,\bar{z}$, which is more complicated. Indeed, it might have additional UV divergences coming from points colliding inside the blob (which can be canceled by adding counter-terms), and also when $z,\bar{z}$ are inside the interval $B$ or near its endpoints. 
For small $z-\bar{z}$ the hypergeometric function in Eq. \eqref{eq:ext_int_hh} has the following expansion
\begin{multline}
\phantom{m}_2 F_1 \l( 1-h_s,h_s,2-h_s,\frac{y-\bar{z}}{z-\bar{z}} \r)=\\- e^{\pm i \pi h_s} \frac{\Gamma(2-h_s) \Gamma(-1+2h_s)}{\Gamma(h_s)} \l( \frac{y-\bar{z}}{z-\bar{z}}\r)^{h_s-1} + \Oc((z-\bar{z})^{h_s}).
\end{multline}
After plugging it in Eq. \eqref{eq:ext_int_hh}, we notice that the factor $(y-\bar{z})^{1-h_s}$ cancels up to a possible phase. This phase is determined by the prefactor $e^{\pm i \pi h}$ which comes from the hypergeometric function branch cut. If $z$ is real and outside the interval $[a,b]$ then the points $y=a$ and $y=b$ come with the same phase and the divergence $(z-\bar{z})^{1-h_s}$ cancels out. However, if $z$ is between $a$ and $b$ then the phase $e^{\pm i \pi h_s}$ is different and the integrand blows up as
\beq
\l( \frac{1}{(z-\bar{z})^{2h_s-1}} \r)^k,
\label{eq:blowup}
\eeq
if $h_s>1/2$, otherwise it is finite.
Does this behavior lead to a divergent integral? Let us assume that $V(z,\bar{z})$ represents a relevant perturbation of the theory, influencing its behavior at long distances, but insensitive to its UV details. Then conformal dimension must be less than $2$, but it should contain at least $k$ insertions of $\Oc_{h_s/2,h_s/2}$, hence 
\beq
k h_s < 2.
\eeq
At the same time $k \ge 1$. Therefore, we can conclude that the negative power of $(z-\bar{z})$ in the integrand is
\beq
k(2h_s-1) = 2 k h_s - k \le 4 -1 =3.
\eeq
So, in principle, the integral in $z,\bar{z}$ can still be divergent and to ensure its convergence we need to impose
\beq
2 k h_s - k \le 1  \ra h_s \le \oh + \frac{1}{2k}.
\eeq
If this condition is satisfied, the divergent contributions only arise from the free-field behavior, hence the ratio (\ref{eq:ratio2}) will be finite. We should emphasize that $k$ here is not a fixed number, but it represents how many operators $\Oc_s$ \textit{could} couple to the interaction vertex. For example, for a scalar field $\phi$, if $\Oc_s = \phi$ and $V = \phi^4$, then $k$ can be $1,2,3$ or $4$. The inequality is the strictest for the maximal possible $k$.

The case of a vector operator with weights $(h_v/2+1,h_v/2)+(h_v/2,h_v/2+1)$ is similar. Instead of Eq. (\ref{eq:scalar_int}), the integrand is $1/((y-z)^{2} |y-z|^{2h_v})$. After integrating over $y$, the resulting expression has the following behavior for small $z-\bar{z}$:
\beq
\frac{1}{(z-\bar{z})^{2 h_v}}.
\eeq
Again, assuming that the interaction vertex has degree $k$, we need to impose the following inequality to ensure that the $z,\bar{z}$ integral is convergent
\beq
h_v \le \frac{1}{2k}.
\eeq
To summarize, we have shown that by using observable-projected ensembles, it is possible to extract UV-finite quantities also in the presence of interactions generated by operators with low scaling dimension.
\section{Numerical Checks}
\label{sec:num_checks}
\begin{figure}[ht!]
\centering
    {\includegraphics[width=0.49\textwidth]{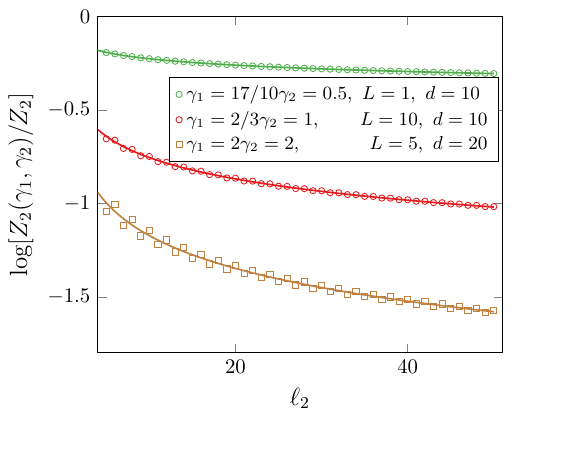}}
    {\includegraphics[width=0.49\textwidth]{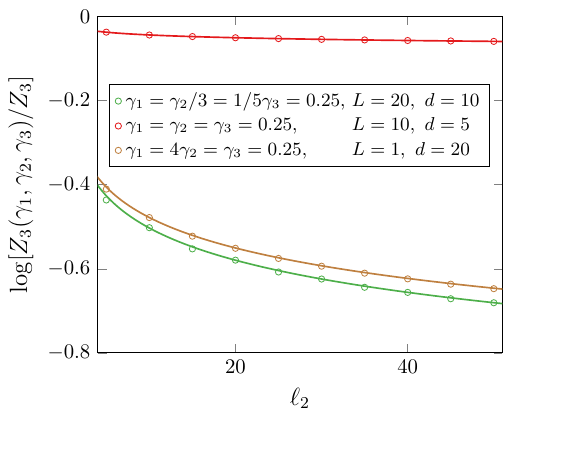}}
     \caption{Normalized (logarithm of the) charged moments defined in Eq.~\eqref{eq:def} as a function of $\ell_2$ for different values of $\gamma_1,\gamma_2, d ,\ell_1$. Here we are considering the ground state of the Hamiltonian \eqref{eq:Hamiltonian} with $\kappa=h=0$.
     The curves correspond to our result~\eqref{eq:traces} with an additive constant obtained from a fit, while the symbols are the exact numerical values from Eq.~\eqref{eq:numerics}. We report only the real part, the imaginary part is due to how the charge is discretized on the lattice. }
     \label{fig:xx}
\end{figure}

In the sections above we have shown it is possible to compute the entanglement properties of the observable-projected ensembles in field theory. Now, we show how it is possible to address this question also numerically, when the observable we measure is the charge. We focus on the ground state of the Hamiltonian
\begin{equation}\label{eq:Hamiltonian}
H=-\frac{1}{2}\sum_{j=-\infty}^{\infty}\left(c^{\dagger}_j c_{j+1}+\kappa c^{\dagger}_jc^{\dagger}_{j+1} +\mathrm{h.c.}+2h c^{\dagger}_jc_j\right),
\end{equation}
for two sets of parameters: when $\kappa=h=1$, it reduces to a critical Majorana chain, while for $\kappa=h=0$ is corresponds to a tight-binding model with $U(1)$ symmetry (whose continuum limit is described in section~\ref{sec:freecompact} for $K=1$). Here $\boldsymbol{c}_j=(c_j^\dagger, c_j)$ are the fermionic operators. We consider an infinite system $A\cup B \cup C$ and we define $\rho_{AB}=\mathrm{Tr}(\rho_{ABC})$ (i.e. $\bar{A}=B\cup C$). The reduced density matrix $\rho_{AB}$ of the ground state of the Hamiltonian~\eqref{eq:Hamiltonian} is
a Gaussian operator in terms of $\boldsymbol{c}_j$~\cite{Peschel_2003}. 
Therefore, we can express our quantities of interest in terms of the 
two-point correlation matrix 
\begin{equation}\label{eq:corr} \Gamma_{jj'}=2\mathrm{Tr}\left[\rho_A\boldsymbol{c}_j^\dagger
 \boldsymbol{c}_{j'}\right]-\delta_{jj'},
\end{equation}
with $j,j'\in AB$. 
Since $\rho_{AB}$ is a Gaussian operator, we can write it as
\begin{equation}\label{eq:ptrace}
    \mathrm{Tr}_B( \rho_{AB}e^{i\gamma Q_B})=\mathrm{Tr}_B\left(\frac{1}{Z_{AB}}e^{-\sum_{j j'}\boldsymbol{c}_j^{\dagger}h_{jj'}\boldsymbol{c}_{j'}}e^{i\gamma\sum_{j j'}\boldsymbol{c}_j^{\dagger}n^B_{jj'}\boldsymbol{c}_{j'}}\right),
\end{equation}
where $n^B_{jj'}=0$ if $j,j' \in A$ and $n^B_{jj'}=\sigma^z\delta_{jj'}$ otherwise, and $Z_{AB}$ is a normalization factor.
The operator above involves the product of Gaussian operators, so it is still a Gaussian operator and by using the Baker-Campbell-Haussdorf formula, we can find~\cite{Fagotti_2010,Ares_2023}
\begin{equation}\label{eq:prod_gaussian}
e^{\sum_{j,j'} \boldsymbol{c}_j^\dagger A_{jj'} \boldsymbol{c}_{j'}}
e^{\sum_{j,j'} \boldsymbol{c}_j^\dagger B_{jj'} \boldsymbol{c}_{j'}}=
e^{\sum_{j,j'} \boldsymbol{c}_j^\dagger H_{jj'} \boldsymbol{c}_{j'}}
\end{equation}
where $H=\log(e^A e^B)$. Therefore, Eq.~\eqref{eq:ptrace} can be written as 
\begin{equation}
    \mathrm{Tr}_B\left(\frac{1}{Z_{AB}}e^{-\sum_{j j'}\boldsymbol{c}_j^{\dagger}h^{\gamma}_{jj'}\boldsymbol{c}_{j'}}\right), \qquad h^\gamma=\log(e^{-h}e^{i\gamma n^B}).
\end{equation}
Moreover, the single-particle entanglement Hamiltonian $h$ of Eq.~\eqref{eq:ptrace} can be expressed in terms of the correlation matrix~\eqref{eq:corr} as 
\begin{equation}\label{eq:single_particle}
    e^{-h}=\frac{1-\Gamma}{1+\Gamma}.
\end{equation}
We can now define a normalized operator with trace 1
\begin{equation}
    \mathrm{Tr}_B\left(\frac{1}{Z^{\gamma}_{AB}}e^{-\sum_{j j'}\boldsymbol{c}_j^{\dagger}h^{\gamma}_{jj'}\boldsymbol{c}_{j'}}\right), \qquad Z^{\gamma}_{AB}=\mathrm{Tr}\left(e^{-\sum_{j j'}\boldsymbol{c}_j^{\dagger}h^{\gamma}_{jj'}\boldsymbol{c}_{j'}}\right),
\end{equation}
and rewrite it as
\begin{equation}
    \mathrm{Tr}_B\left(\frac{1}{Z^{\alpha}_{AB}}e^{-\sum_{j j'}\boldsymbol{c}_j^{\dagger}h^{\gamma}_{jj'}\boldsymbol{c}_{j'}}\right)=\frac{1}{Z^{\gamma}_{A}}e^{-\sum_{j j'}\boldsymbol{c}_j^{\dagger}h^{\gamma,A}_{jj'}\boldsymbol{c}_{j'}}, \qquad Z^{\gamma}_{A}=\mathrm{Tr}\left(e^{-\sum_{j j'}\boldsymbol{c}_j^{\dagger}h^{\gamma,A}_{jj'}\boldsymbol{c}_{j'}}\right),
\end{equation}
with 
\begin{equation}
    e^{-h^{\gamma}}=\frac{1-\Gamma^{\gamma}}{1+\Gamma^{\gamma}}.
\end{equation}
For a general 
non-Hermitian matrix $H$, we can use that~\cite{Fagotti_2010}
\begin{equation}\label{eq:logtr}
    \mathrm{Tr}(e^{\sum_{j,j'} \boldsymbol{c}_j^\dagger H_{j j'}
    \boldsymbol{c}_{j}})=\sqrt{\mathrm{det}(1+e^{H})},
\end{equation}
and hence
\begin{equation}
Z^\alpha_{AB}=\det\left(1+\frac{1-\Gamma}{1+\Gamma}e^{i\gamma n_B}\right),\qquad Z^\alpha_{A}=\det\left(1+\frac{1-\Gamma^{\gamma}_A}{1+\Gamma^{\gamma}_A}\right).
\end{equation}
We remark that $\Gamma^{\gamma}_A$ can be obtained by restricting $\Gamma^{\gamma}$ to the subsystem $A$. If we put everything together, we find that 
\begin{equation}
    \mathrm{Tr}_B\left(\frac{1}{Z_{AB}}e^{-\sum_{j j'} \boldsymbol{c}_j^{\dagger}h_{jj'} \boldsymbol{c}_{j'}}e^{i\gamma\sum_{j j'} \boldsymbol{c}_j^{\dagger}n^B_{jj'} \boldsymbol{c}_{j'}}\right)=\frac{Z_{AB}^{\gamma}}{Z_{AB}}\frac{1}{Z_A^{\gamma}}e^{-\sum_{j j'} \boldsymbol{c}_j^{\dagger}h^{\gamma,A}_{jj'} \boldsymbol{c}_{j'}},
\end{equation}
and for a generic R\'enyi index $n$
\begin{equation}\label{eq:numerics}
    \mathrm{Tr}_A\left[\prod_{j=1}^n\left(\frac{1}{Z_{AB}}e^{-\sum_{j j'}\boldsymbol{c}_j^{\dagger}h_{jj'}\boldsymbol{c}_{j'}}e^{i\gamma_j\sum_{j j'}\boldsymbol{c}_j^{\dagger}n^B_{jj'}\boldsymbol{c}_{j'}}\right)\right]=\frac{1}{Z_{AB}^n}\prod_{j=1}^n\left[\frac{Z_{AB}^{\gamma_j}}{Z_A^{\gamma_j}}\right]\det \left(1+\prod_{j=1}^n\frac{1-\Gamma^{\gamma_j}_A}{1+\Gamma^{\gamma_j}_A}\right).
\end{equation}
If we denote by
\begin{equation}\label{eq:def}
    \frac{Z_n(\gamma_1,\dots,\gamma_n)}{Z_n}=\frac{\mathrm{Tr}_A[\prod_{j=1}^n\mathrm{Tr}_B( \rho_{AB}e^{i\gamma_j Q_B})]}{\mathrm{Tr}(\rho_A^n)},
\end{equation}
we can cross-check the result in Eq.~\eqref{eq:traces} by choosing $\kappa=h=0$ in Eq.~\eqref{eq:Hamiltonian}. We report the comparison between the exact numerical results and our analytical predictions in Fig.~\ref{fig:xx} for fixed subsystem size $L$, $a=L+d$, and by varying $b-a\equiv \ell_2$.

\begin{figure}[ht!]
\centering
  {\includegraphics[width=0.495\textwidth]{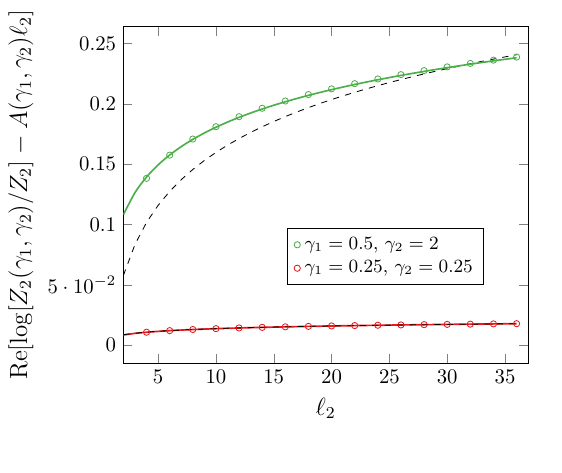}}
    {\includegraphics[width=0.495\textwidth]{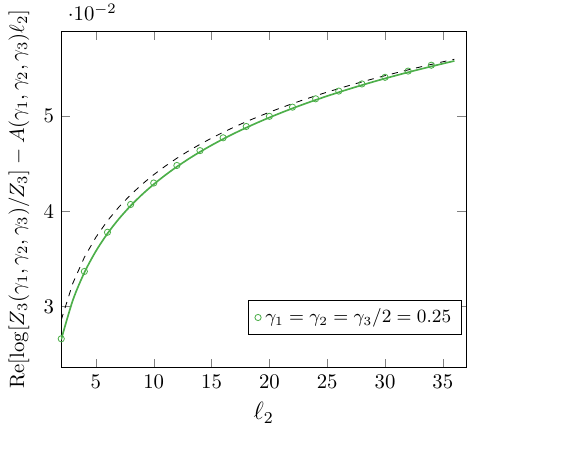}}
    %{\includegraphics[width=0.495\textwidth]{numerics/chargedmoments_Isingn3_b.pdf}}
     \caption{Real part of the normalized (logarithm of the) charged moments defined in Eq.~\eqref{eq:def} as a function of $\ell_2$ for different values of $\gamma $, $\ell_1=d=10$ for the ground state of the Hamiltonian \eqref{eq:Hamiltonian} with $\kappa=h=1$. Once we subtract the non-universal contribution (which is linear in $\ell_2$ and we denote it by $A(\gamma_i)\ell_2$), the solid curves have been obtained by plugging Eq.~\eqref{eq:ising} in~\eqref{eq:gaussian_gen} with $\gamma_i\to \frac{1}{\pi}\textrm{arctanh}(\tan(\gamma_i/2))$ and an additive fitted constant. The symbols are the exact numerical values from Eq.~\eqref{eq:numerics}. The black dashed lines correspond to the quadratic approximation in $\gamma_i$.}
     \label{fig:ising}
\end{figure}

We repeat the same analysis for the ground state of the Majorana chain at its critical point by setting $\kappa=h=1$ in Eq.~\eqref{eq:Hamiltonian}. Despite the lattice operator is quadratic, $Q=\sum_jc^{\dagger}_j c_j$, as well as its continuum counterpart (one of the primary fields of the Ising CFT with scaling dimension $1$), it does not satisfy Wick's theorem, in the sense that a simplification like the one in Eq. \eqref{eq:central_Q} does not hold \footnote{We remark that the fermionic representation of this operator, $\varepsilon=\bar{\psi}\psi$, permits to easily compute all its correlators using Wick’s theorem. However, the operator $\varepsilon$ itself does not satisfy the Wick theorem unless the correlators are evaluated along the real line. However, in general, the regularization scheme used in Eqs. \eqref{eq:div1}, \eqref{eq:conf_int} shows that we need to evaluate correlators defined on the full complex plane to avoid divergences, and only after the integration we can send $\varepsilon
\to 0$. Thus, Wick's theorem for $\varepsilon$ itself does not hold.}. Therefore, we can use our result~\eqref{eq:ising} only as a perturbative expansion in $\gamma$.
From Eq.~\eqref{eq:ising} we expect the presence of a linear term in $\ell_2$, whose prefactor $1/\epsilon$ is non-universal, and a subleading logarithmic growth. 
To find the exact result in all orders in $\gamma$, we should map Eq.~\eqref{eq:def} to a cylinder with $n$ defects, rather than to the complex plane, as we did in Eq.~\eqref{eq:map_gen}. 
After the mapping, Ref.~\cite{Fossati_2024} proved that this quantity can be related to the ground state energy of the massless Majorana fermion theory on a circle with
marginal point defects, at least in the limit in which $a\to 0, b\to L$, i.e. $A$ and $B$ coincide. Our case is slightly different because the boundary conditions that the fields obey along $B$ on the Riemann surface appearing in Fig.~\ref{fig:cft_sheets} are simpler, so each $i$-th sheet can be related to the massless Majorana fermion theory with only one single defect, that in our original problem would be $e^{i\gamma_i Q_B}$. We can borrow the results from Ref.~\cite{Fossati_2024} and we find that our results match up to rescaling $\gamma_i\to \textrm{arctanh}(\tan(\gamma_i/2))$ in Eq.~\eqref{eq:gaussian_gen}.
This rescaling was derived by mapping the system to a cylinder but it is also possible to obtain it from the direct perturbative expansion in $\gamma_i$ using the known correlation functions of $\varepsilon$. We have examined perturbatively the subleading (quartic) term in $\gamma_i$ and found 
\footnote{This computation involves computing a four-fold integral which is hard to evaluate. However, we are only interested in the term  $\log(\ell_2/\ep)$. The coefficient in front of this term can be found by numerically computing the integral for complex values of $\ep$ and extracting the branch-cut of the logarithm. 
}
that it matches the expansion of $\textrm{arctanh}(\tan(\gamma_i/2))$. 
We cross-check this with the exact numerical results in Fig.~\ref{fig:ising} focusing on the universal logarithmic term, and we remark that for small values of $\gamma_i$, $ \textrm{arctanh}(\tan(\gamma_i/2))\simeq \gamma_i/2$ and our quadratic approximation is valid, as we would expect (dashed black lines). In the large $\ell_2$-limit, we stress that this approximation is enough to get the result in Eq.~\eqref{eq:gauss_gen2}.

%\begin{figure}[ht!]
%\centering
%\includegraphics[width=0.32\textwidth]{numerics/chargedmoments_vsl1n3.pdf}
     %\includegraphics[width=0.32\textwidth]{numerics/chargedmoments_vsl1n3_2.pdf}
   %  \caption{Normalized charged moments defined in Eq.~\eqref{eq:def} as a function of $\ell_1$. The smooth curves are Alexey's prediction $\frac{1}{4\pi^2}\sum_{jk}\alpha_j\alpha_k M_jk$ with an additive constant obtained from the fit, while the oscillating lines are the exact numerical values. The oscillations are due to non-universal contributions and in general the dependence on $\ell_1$ is almost a constant function.}
 %    \label{fig:l1}
%\end{figure}
\section{Conclusions}
In this manuscript, we have investigated the projected ensembles from a novel perspective, considering set of states generated not by partial projective measurements on the system but by measurements of a given observable, like the number of particles or the magnetization. After this operation, we have considered an ensemble of mixed-states and we have investigated the entanglement properties, both for any individual outcome and the measurement-induced entanglement (cf. Eq.~\eqref{eq:SIE}). The advantages of considering an observable-projected ensemble are two-fold: as we have shown in the main text, it is amenable to an analytical solution, but at the same time these properties can also be experimentally probed, by using classical shadows and randomized measurements.
For a free compact boson, we could compute explicit analytical expressions that have also provided an upper bound on the amount of the information that one can extract about the subsystem $A$ after the measurement of the charge in a non-complementary region, in particular how this is related to the size of the measured system. For generic Gaussian theories, similar computations can be performed, even though the dependence on the geometry becomes more involved. Finally, we suggest how our observable-projected ensemble can be used to extract universal quantities, by properly regularizing UV divergences.

There are several future directions that our manuscript might open. For usual projected ensembles, after averaging the $k$-moments, i.e. constructing
$
\Ec_k = \mathbb{E}_\Ec \l( | \Psi_i \ket \bra \Psi_i |\r)^{\otimes k},$
one can show that, in the absence of symmetries and conservation laws, ergodic systems produce subsystems in the maximally mixed states~\cite{cotler23,choi23}. We have already mentioned in the introduction that these moments are close to 
%The corresponding counterpart of this statement for $\rho_k$ is that $\rho_k$ is close 
an ensemble of Haar-random states.
%: for an appropriate choice of norm $||\cdot||$ (e.g. Schatten 1-norm),
%\beq
%\label{eq:k_moment}
%|| \rho_k - \rho_{\rm Haar, k} || < \ep,
%\eeq
%where  $\rho_{\rm Haar, k } = \int dU_A ( U_A^{\dagger} |\psi_A \ket \bra \psi_A| )^{\otimes k}$ and $\epsilon$ is an arbitrary small number.
%The question whether Eq. (\ref{eq:k_moment}) holds is known as \textit{deep thermalization.} By now there exists a large body of literature addressing this question in various discrete systems. In many cases there is a time-scale separation, when it takes longer times to reach higher $k$ designs.
We can ask a similar question for observable-projected ensembles: after preparing a given state, $\rho$, and letting it evolve with Haar random dynamics, we can construct the
ensemble~\eqref{eq:mathEO} and ask whether the corresponding $k$-moments are close to some universal distribution and they also satisfy a maximum entropy principle~\cite{federica}. To address this question we can also start from an easier setup in which $A$ and $B$ are complementary systems, we fix the subsystem size of $A$ and we scale the subsystem size of $B$.
In general, it would be interesting to study the observable-projected ensemble in a dynamical setup and investigate if phenomena like deep thermalization also occur in this case.

In section~\ref{sec:chargeproj} we have outlined an experimental proposal to probe the properties of the charge-projected ensembles. This can be doable, for instance, by exploiting the preparation of the ground state of the XXZ spin chain (recently discussed in~\cite{Najafi2024}), which is a microscopic realization of the compact boson studied in section~\ref{sec:freecompact}.

Our work provides a strategy to study projected ensembles in a field theory setup, by implementing them through local operators depending on the measured observable. On the way to study also the projected ensembles~\eqref{eq:proj_ens} in field theory, Ref.~\cite{hoshino2024} points out that performing a Bell measurement amounts to inserting a boundary changing operator between the measured and the unmeasured part. This implies that if the boundary condition is conformal invariant, the measurement-induced entanglement might reduce to a correlation function of boundary changing operators. A thorough analysis in this direction might allow for studying the projected ensembles in conformal field theories. 

Finally, it would be interesting to use observable-projected ensembles to extract information about the initial wavefunction.
For instance, in~\cite{Lin2023probingsign}, the authors found that measuring stabilizer sign-free states in the sign-free basis cannot generate more correlations than those that already exist, by providing some bounds on the MIE for different systems.  
On the other hand, these bounds are violated for generic (non-stabilizer) states, for example, critical states with sign structure.

\section*{Acknowledgments}

\noindent We would like to thank Filiberto Ares, Andreas Elben, Giuseppe Di Giulio, Hsin-Yuan Huang, Lorenzo Piroli, John~Preskill, Sridip Pal, Federica Surace, and Vittorio Vitale for useful discussions. We also thank Wen Wei Ho for suggesting potential future studies about the observable-projected ensembles.
SM thanks the support from the Caltech Institute for Quantum Information and Matter and the Walter Burke Institute for Theoretical Physics at Caltech. 
AM acknowledges funding provided by the Simons Foundation, the DOE QuantISED program (DE-SC0018407), and the Air Force Office of Scientific Research (FA9550-19-1-0360). The Institute for Quantum Information and Matter is an NSF Physics Frontiers Center. AM was also supported by the Simons Foundation under grant 376205.

\bibliographystyle{quantum}
\bibliography{refs}
\end{document}